\documentclass[review,number,sort&compress]{elsarticle}

\pdfoutput=1

\usepackage{graphicx}
\usepackage{epstopdf}
\usepackage{subfigure}
\usepackage{amsmath}
\usepackage{amssymb}
\usepackage{afterpage}
\usepackage{graphics}
\usepackage{float}
\usepackage{rotating}
\usepackage{xspace}
\usepackage{dcolumn}
\usepackage{bm} 
\usepackage{lineno}
\usepackage{placeins}
\usepackage{changepage}

\journal{Nucl.Inst.Meth.A}

\begin{document}


\begin{frontmatter}



\title{A Novel Use of Light Guides and Wavelength Shifting Plates for the Detection of Scintillation Photons in Large Liquid Argon Detectors}
\tnotetext[t1]{FERMILAB-PUB-17-488-ND-PPD, arXiv:1710.11233 [physics.ins-det]}

\author[label1]{B.~Howard}
\address[label1]{Indiana University, Bloomington, Indiana 47405, USA}

\author[label1]{S.~Mufson}


\author[label1]{D.~Whittington\fnref{jose}}\fntext[jose]{currently: Syracuse University, Syracuse, NY, 13244, USA}

\author[label1]{B.~Adams}

\author[label1]{B.~Baugh}

\author[label1]{J.R.~Jordan\fnref{maria}}\fntext[maria]{currently: University of Michigan, Ann Arbor, MI, 48109, USA}

\author[label1]{J.~Karty}

\author[label1]{C.T.~Macias}

\author[label3]{A.~Pla-Dalmau }
\address[label3]{Fermi National Accelerator Laboratory, Batavia, Illinois 60510, USA}

\date{\today}          

\begin{abstract}
Scintillation light generated as charged particles traverse large liquid argon detectors adds valuable information to studies of weakly-interacting particles.  
This paper uses both laboratory measurements and cosmic ray data from the Blanche dewar facility at Fermilab to characterize the efficiency of the photon detector technology developed at Indiana University for the single phase far detector of DUNE.  
The efficiency of this technology was found to be 0.48\% at the readout end when the detector components were characterized with laboratory measurements. 
A second determination of the efficiency using cosmic ray tracks is in reasonable agreement with the laboratory determination.  
The agreement of these two efficiency determinations supports the result that minimum ionizing muons generate ${\mathcal N}_{phot} = 40,000$ photons/MeV as they cross the LAr volume.

\end{abstract}

\end{frontmatter}

\hfill
\newpage


\section{Introduction}
\label{sect:intro}

The analysis of the scintillation light generated as charged particles traverse large liquid argon (LAr) time-projection chamber (TPC) detectors adds valuable information to studies of weakly-interacting particles.  Most importantly, the leading edge of the scintillation light pulse yields sub-mm accuracy in reconstructing the absolute position of the event in the drift direction~\cite{bib:DUNE-CDR-vol4}.  In addition, the scintillation light can provide a trigger for proton decay and supernova neutrinos, as well as a handle useful in the rejection of uncorrelated cosmic backgrounds~\cite{bib:DUNE-CDR-vol4}.  Further, scintillation light can be used as a tool for particle identification~\cite{bib:pulseShape2}.

In this paper we describe a prototype technology designed to detect the LAr scintillation photons in the single phase far detector of the Deep Underground Neutrino Experiment (DUNE)~~\cite{bib:DUNE-CDR-vol1}.  Since significant photocathode coverage in a large volume detector like DUNE is prohibitively expensive, this technology uses light guides to collect and channel the scintillation photons to a small number of photosensors at their ends.  Although highly efficient light guides are commercially available for collecting and transporting visible photons, scintillation photons from LAr are generated in the vacuum ultraviolet (VUV) at 128 nm.  Accordingly, this technology covers the faces of the commercial light guides with wavelength shifting plates that convert the VUV photons into the visible.   Although its efficiency is relatively  modest, this technology can be made practical because LAr is a copious source of scintillation light, producing tens of thousands of VUV photons per MeV along a track, and pure liquid argon is transparent to its own scintillation light.  

The technology described here is one of multiple photon detector (PD) technologies that will be tested in ProtoDUNE-SP~\cite{bib:protoDUNE}, an experimental program at CERN designed to evaluate the technologies proposed for the single-phase DUNE far detector. 

The primary purpose of this paper is to characterize the prototype PD technology developed at Indiana University (IU) by reporting the results of experimental tests both in the laboratory at IU and at the Blanche LAr test facility at Fermilab.  These tests were mainly intended to validate the design using cosmic ray data and to verify that the quality control (QC) procedures used in detector construction accurately predicted their performance.   In these studies both laboratory measurements and Blanche cosmic ray data are used to  determine the absolute efficiency of the detectors.  There is one free parameter available to match these two independent determinations of the efficiency -- the light yield of cosmic muons as they traverse LAr.  In most simulations, minimum ionizing muons are assumed to generate $\sim$40,000 photons/MeV as they cross the LAr volume~\cite{bib:Miyajima,bib:Doke1,bib:Doke2,bib:scintYield2}.   Our Blanche experiment provides a valuable data set to cross check this expectation.  

\section{Photon Detector Design}
\label{sec:PhotonDetectorDesign}

\begin{figure}[h]
\centerline{\includegraphics[width=.95\textwidth]{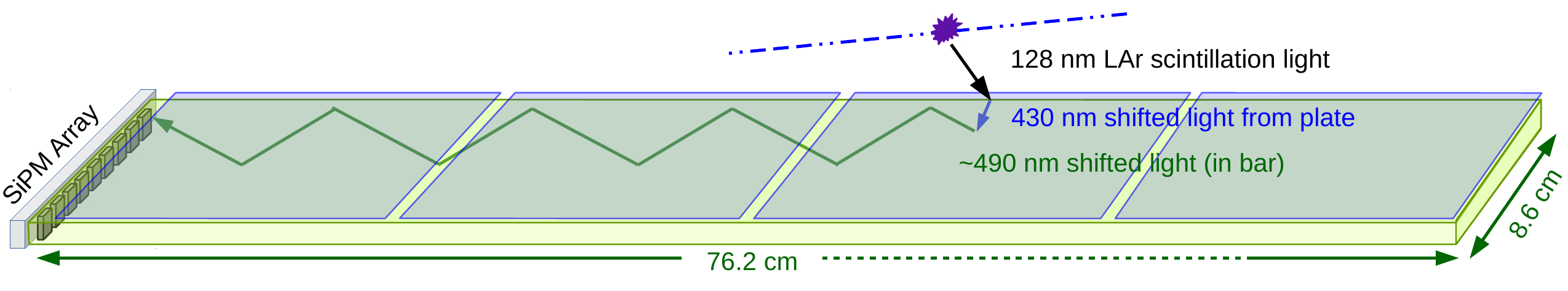}}
\caption{The design concept for the photon detector technology.  VUV scintillation photons at 128~nm from LAr strike wavelength shifting plates embedded with TPB where they are converted to $\sim$430~nm photons.  These visible photons are subsequently absorbed by a commercial light guide where they are again wavelength shifted to $\sim$490~nm photons and channeled to silicon photomultipliers (SiPMs) at the end.  The dimensions shown are for the photon detectors used in the Blanche dewar facility where 4 plates are used.}
\label{fig:PDcartoon} 
\end{figure}

The design concept for the photon detector technology described in this paper is shown in Fig.~\ref{fig:PDcartoon}.  Scintillation photons from liquid argon at 128~nm strike one of four acrylic plates with TPB (1,1,4,4-tetraphenyl-1,3-butadiene) embedded in their surfaces.  The TPB in the struck plate converts VUV scintillation photons to visible photons typically in the range 420 -- 450~nm.  These  photons are transmitted through the plate and are subsequently absorbed by a commercial light guide made by Eljen Technology\footnote[1]{http://www.eljentechnology.com}.  In the light guide, photons are again wavelength shifted to the range 480 -- 510~nm and transported to silicon photomultipliers (SiPMs) at the end.  This design supports a maximum of 12 $6\times6$~mm$^2$ SensL C-series SiPMs\footnote[2]{http://sensl.com/products/c-series/}.  For the experiments described here, 8 $6\times6$~mm$^2$ SiPMs were used to read out the light guide.  

\begin{figure}[h]
\centering
\includegraphics[width=.75\textwidth]{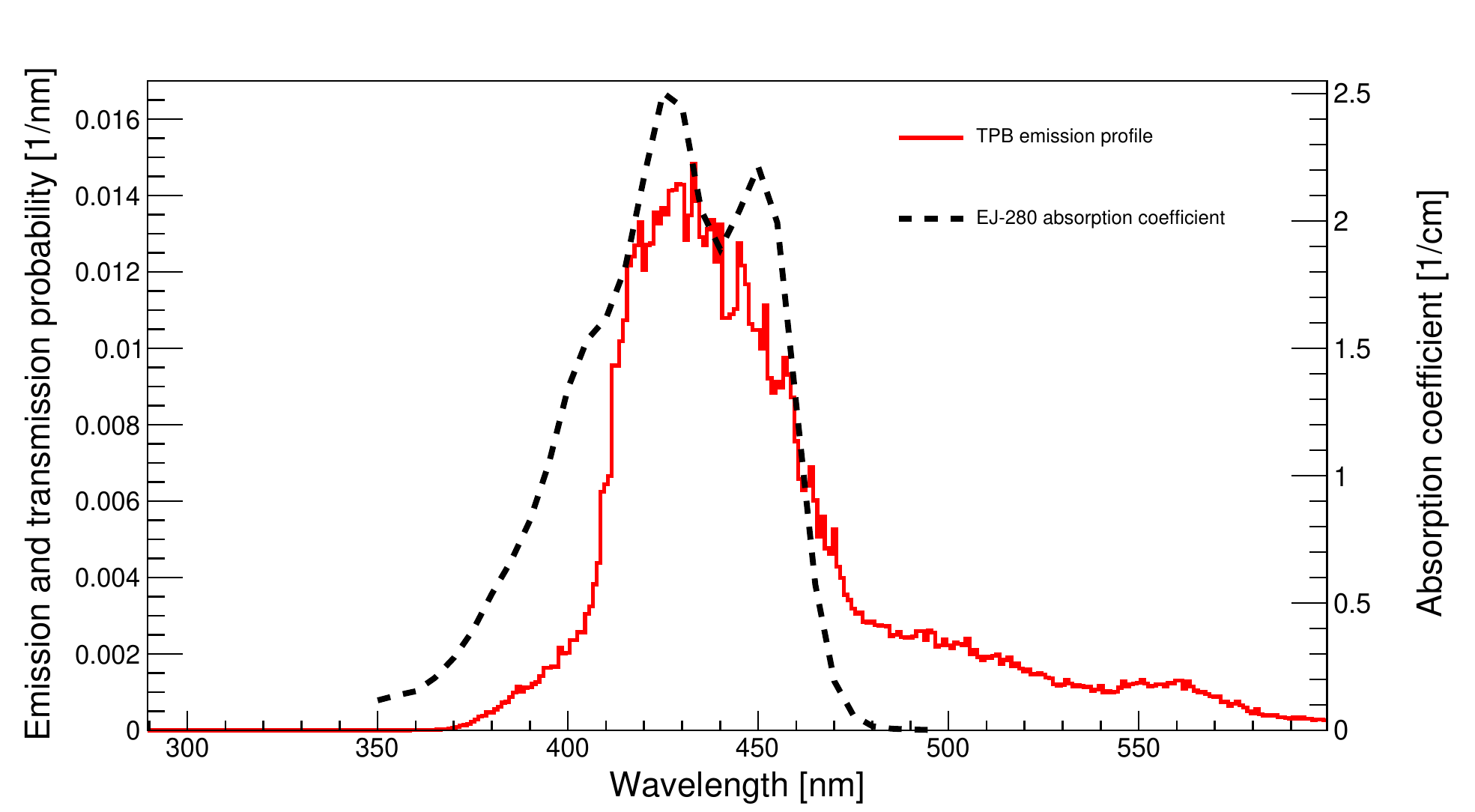}
\includegraphics[width=.75\textwidth]{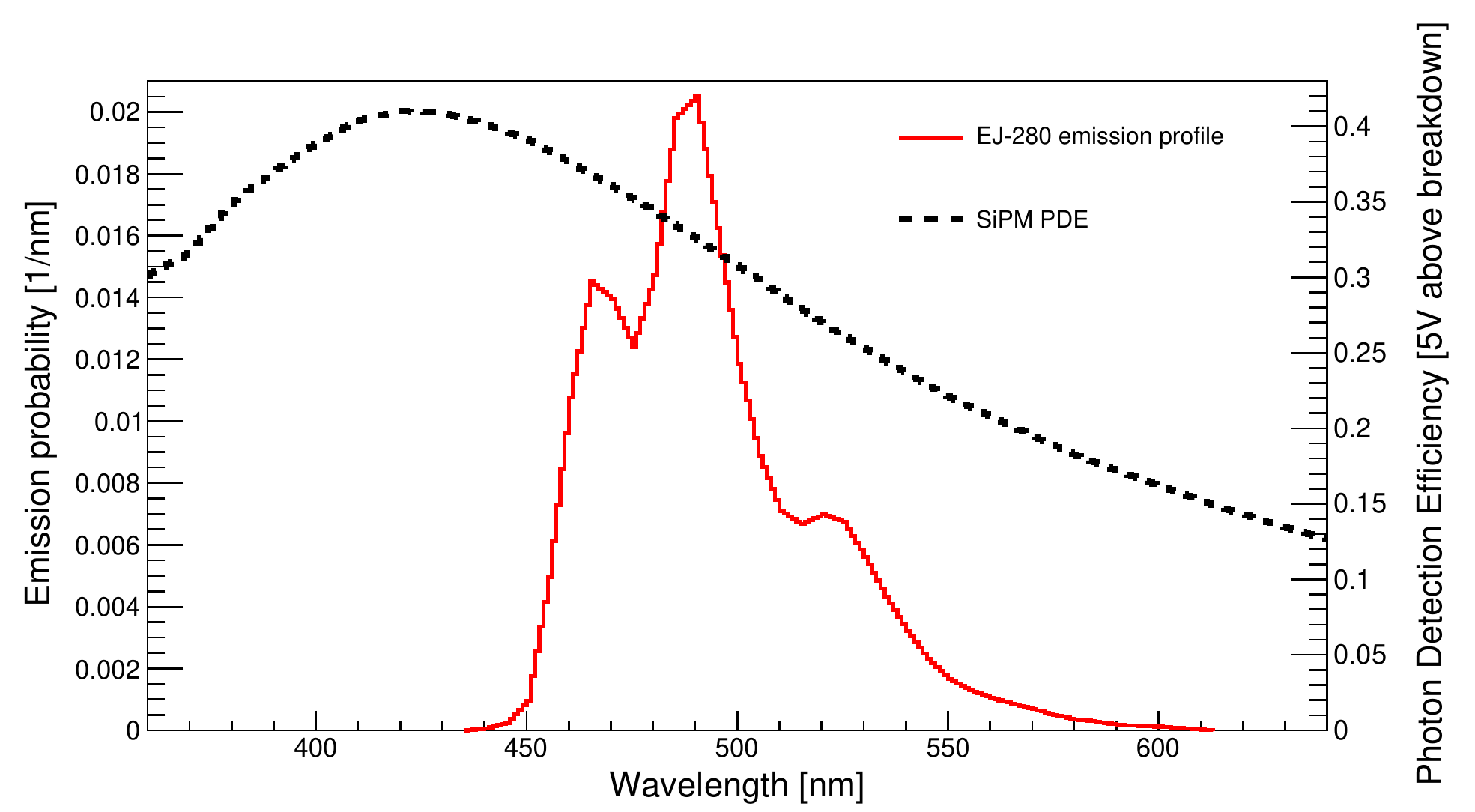}
\caption{{\it top panel}: the TPB emission spectrum for the wavelength shifting plates, calculated as the product of the TPB emission profile normalized to unit area and the transmission function through 1/16$''$ thick acrylic plates, and the absorption spectrum for the EJ-280 Eljen light guides; {\it bottom panel}: the emission spectrum for the EJ-280 Eljen light guides normalized to unit area and the PDE for the SensL C series SiPMs at 5 V overvoltage.}
\label{fig:PDsummary} 
\end{figure}

In Fig.~\ref{fig:PDsummary} this photon detector technology is characterized in more detail.  The top panel shows the TPB emission spectrum for the wavelength shifting plates that absorb the 128~nm scintillation photons and then re-emit $\sim$430~nm visible photons.  The emission spectrum has been calculated as the product of the area-normalized TPB emission profile and the transmission function through 1/16$''$ thick acrylic plates.  Also shown is the absorption spectrum for the Eljen EJ-280 light guides that reabsorb these visible photons.  The TPB emission profile was determined by a Hitachi F-4500 fluorescence spectrophotometer.  Pieces cut from three plates were illuminated from the front with 200~nm light to simulate the excitation by 128~nm photon and the wavelength shifted spectrum was measured at 90$^\circ$ from the beam direction.  Gehman {et al.}~\cite{bib:gehman} show that the visible re-emission spectra for TPB from light at 128~nm and 200~nm are virtually indistinguishable.  The emission spectrum used to calculate the profile shown is the average of the area-normalized spectra from the three plate samples.  The transmission function for the visible photons through the $\frac{1}{16}''$ acrylic plate was measured using a Cary 50 UV-VIS spectrophotometer.  The plastic sample was glued to a 1x1 cm plate and held perpendicular to UV-VIS beam in the cuvette holder of the spectrophotometer.  Data were recorded from 600 nm to 250 nm in a 0.5 sec integration.  The background was subtracted from the plastic spectrum automatically while the plastic data were recorded.  The absorption spectrum for the EJ-280 light guides was provided by Eljen\footnotemark[1].  

The bottom panel in Fig.~\ref{fig:PDsummary} shows the area-normalized emission spectrum for the EJ-280 Eljen light guides and the photon detection efficiency (PDE) for the SensL C series SiPMs\footnote[2]{http://sensl.com/products/c-series/}.  The emission spectrum for the Eljen EJ-280 light guides was provided by Eljen\footnotemark[1] and has been normalized to unit area.  The absolute PDE for the SiPMs was provided by SensL\footnotemark[2] at 5~V above breakdown voltage, the operating bias voltage used in this experiment in the Blanche dewar facility.

\section{Laboratory Measurements of the Photon Detector Design}
\label{sect:labMeasurements}

\subsection{SiPM Photodectors}
\label{sect:sipm}

The photodectectors used in this experiment are SensL C-series MicroFC-60035-SMT SiPMs\footnotemark[2].  Their performance at cryogenic temperatures has been described in~\cite{bib:TallBo}.  Each SiPM has an active area of $6 \times 6$ mm$^2$.  They are made up of an array of 18,980 microcell photodiodes, each of which is 35 $\mu$m on a side, and the microcell fill factor on the chip is 64\%. The SiPMs are reverse-biased at 25.5 V.  

Table~\ref{tab:SiPM-Noise} lists the operating characteristics of 7 SensL SiPMs used in the Blanche experiment, as determined by dark tests in liquid nitrogen (LN2)~\cite{bib:TallBo} once the experiment was completed.   (The photon detector paddle ``PD1'' is defined in Fig~\ref{blancheExpt}). The breakdown voltage ($V_b$) given is the y intercept of a straight line fit to dark noise measurements made with SiPM bias voltages ranging from 25.5V - 29.5V in steps of 1~V.  The PDE for the SensL SiPMs shown in the bottom panel of Fig.~\ref{fig:PDsummary} is for an operating voltage of 5 V above the breakdown voltage, which is 25.5 V.  These LN2 tests have also been used to show that the afterpulsing probability was less than 1\% in $\sim10~\mu$s.  After $10~\mu$s any additional afterpulse would be counted as dark noise.  
Although the protype detector analyzed in this experiment was populated with 8 working SiPMs, one of them failed during these LN2 dark tests.
\begin{table}[h]
  \begin{center}
    \caption{Dark noise characteristics of 7 SiPMs on PD1 in LN2 at $V_{\text{b}}$ = 25.5 V.}
    \vspace{0.2em}
    \label{tab:SiPM-Noise}
    \begin{tabular}{| c |  c  c  c  c  |}
      \hline
      \hline
      SiPM  & Noise & ~~~Cross-Talk  & Gain & Break.Volt. \\
            &    [Hz]    & ~~~Probability  &  &   [V]    \\  
      \hline
      
      0&  14 &  0.20 & ~~4.6$\times10^6$~~ & 20.5  \\
      1 &  11 &  0.19 & 4.5$\times10^6$ & 20.6  \\
      2 &  13 & 0.19 & 4.6$\times10^6$ & 20.5  \\
      3 &  13 &  0.20 & 4.6$\times10^6$ & 20.5  \\
      4 &  14 & 0.19 & 4.6$\times10^6$ & 20.4  \\
      5 &  11 &  0.19 & 4.5$\times10^6$ & 20.6  \\
      6 &  14 &  0.19 & 4.5$\times10^6$ & 20.6  \\
       \hline
      \hline
      mean  & 13   & 0.19 & 4.6$\times10^6$ & 20.5 \\
        \hline
      \hline
   \end{tabular}
  \end{center}
\end{table}

\subsection{Wavelength Shifting Plates}
\label{sect:plates}

\subsubsection{Wavelength Shifting Plate Production}
\label{sect:plateProduction}

The wavelength shifting plates used in the Blanche experiment were cut into 9$\frac{1}{2}''$ lengths from 3$''$ wide $\times$ $\frac{1}{16}''$ thick acrylic sheets purchased from McMaster-Carr\footnote[3]{https://www.mcmaster.com}.  The plates were coated with a mixture of 5 g of TPB to 1,000 g dichloromethane (DCM).  DCM is a common solvent that readily dissolves the TPB.  The  “scintillation” grade ($\geq$ 99\% pure) TPB was purchased from Sigma-Aldrich\footnote[4]{http://www.sigmaaldrich.com}.  The plates were coated with TPB using a HVLP (High Volume Low Pressure) sprayer system under a fume hood.  As a prototype technology, the spraying has not been commercialized and was done by hand to approximate a standard established to have a relatively high VUV photon conversion efficiency.
Once the plates were coated, they were baked overnight in a vacuum oven at 80$^\circ$ C. This temperature is just below the glass transition point of the acrylic and allows the TPB to be incorporated into the surface of the plastic, making it unlikely that the TPB will flake off in LAr.  Over many years of making wavelength shifting plates, TPB was never observed to flake off in the lab or in experiments at Fermilab.

The emission spectrum of the wavelength shifting plates is shown in the top panel of  Fig.~\ref{fig:PDsummary}.  The performance of the wavelength shifting plates depends most importantly on its efficiency in converting 128~nm light into the visible.  A large number of plates were produced whose relative brightness could be compared with one another and then the brightest ones were selected from this group.  The conversion efficiencies for these selected plates were then computed, as described in \S\ref{sect:platePerformance}.

The relative brightnesses of the wavelength shifting plates was evaluated using a commercial McPherson model 234/302 VUV monochromator\footnote[5]{http://mcphersoninc.com} with a $^2$H lamp source.  A schematic diagram of the monochromator can be found on the McPherson website \\(https://mcphersoninc.com/spectrometers/vuvuvvis/model234302.html); \\the sample plate is placed in a custom holder at the exit slit.  A plate is exposed to 128 nm light in the monochromator; the TPB in the plate converts the 128 nm light to the visible; and the wavelength shifted visible light is quantified by the current read out by a SensL SiPM.  During the ongoing evaluations of wavelength shifting plates, the MgF$_2$ window in the monochromator lamp housing gets fogged by material deposited on its surface, which has the effect of reducing the 128~nm intensity falling on the sample.  To monitor this falling lamp output (between window cleanings), the 128~nm light was quantified typically by the current read out by a VUV photodiode both before and after the sample was exposed.  Since it is assumed that the output of the TPB is proportional to the 128~nm exposure, the brightness of the wavelength shifting plates was evaluated by the ratio of the current read out by the SiPM when the plates were exposed to 128~nm light divided by the average current read out at 128~nm by the VUV  photodiode.  Using this ratio, the wavelength shifting plates in the Blanche experiment were selected from 31 made between July 2016 and August 2016.  

Since the monochromator can only accommodate 1$\frac{1}{2}'' \times 1''$ test samples in its vacuum chamber, a plate's brightness is characterized by the average brightnesses of two 1$''$ breakout tabs cut off from either side of the plate once it had been coated.  The manner in which the wavelength shifting plates are manufactured raises two issues.  First, how uniform is the coating?  Several studies in which plates were cut in pieces and tested in the VUV monochromator show that the variations in coating efficiency typically fall in the range 5-15\%.  As shown below, the plates used in the Blanche experiment fall in this range.  Second, how representative of the average brightness are the two breakout tabs on either side of the plate?  Studies of the brightness of the breakout tabs compared with the average brightness of the cut pieces from several plates show a wider range of variability, typically 5-20\%.  

Since the monochromator measurements of the plates have been made at room temperature, plates were tested in LAr to determine whether their relative brightnesses at room temperature predicted their behavior in LAr.  The results of tests similar to those described in \S\ref{sect:lightguides} clearly demonstrated that plates which were brighter at room temperature in the monochromator were also brighter when exposed to scintillation light from an $^{241}$Am source in LAr.  
\begin{figure}[h]
\centerline{\includegraphics[width=.85\textwidth]{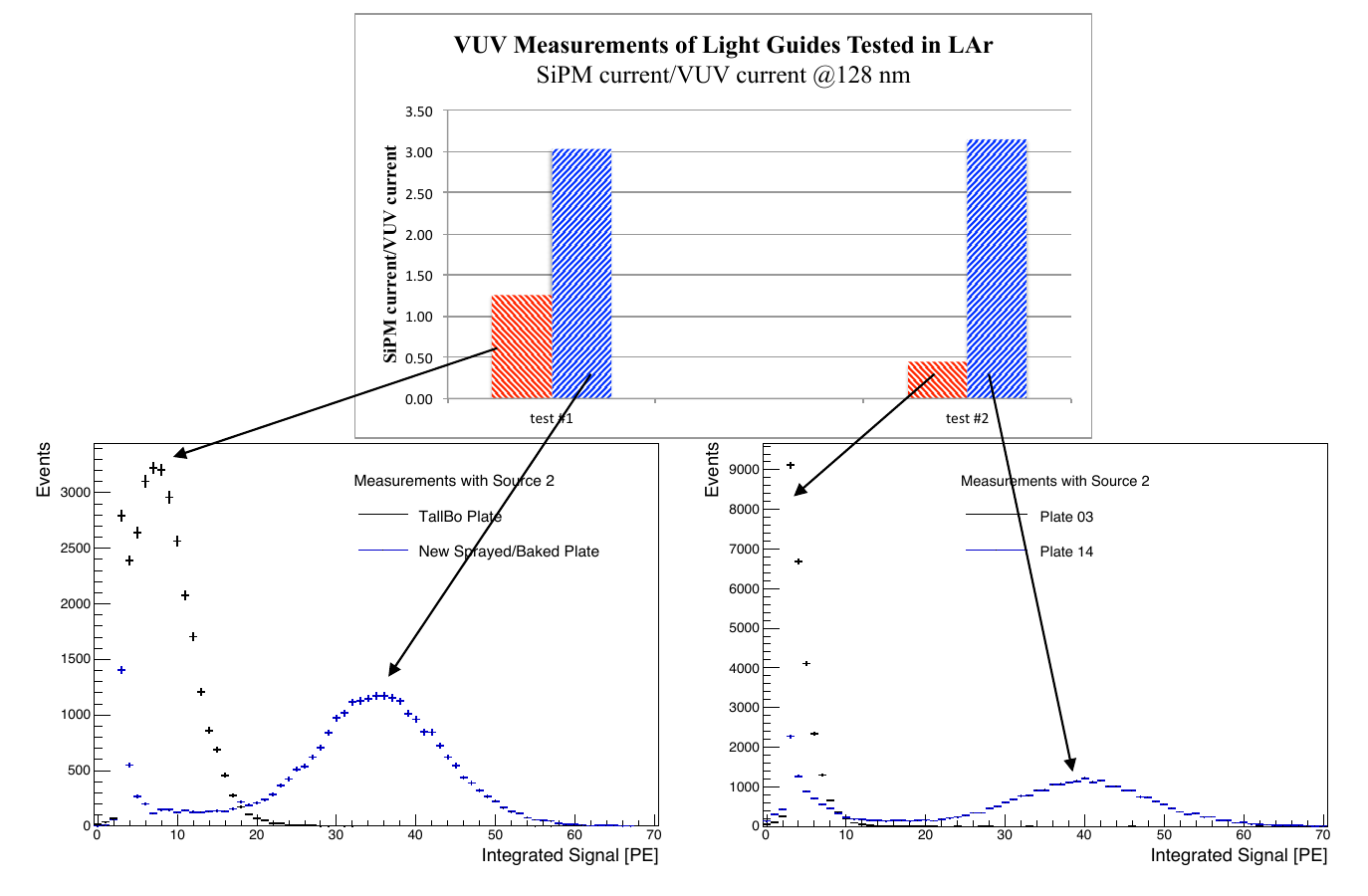}}
\caption{Comparison of the brightness of two sets of wavelength shifting plates in the VUV monochromator and in LAr.  The plates tested were of the same design as used in Blanche but from previous experiments such as in the TallBo dewar facility at Fermilab.  The top bar graphs show the relative brightness differences of the plates in the VUV monochromator.  In LAr the alpha peaks for the brighter plates in the monochromator are significantly brighter than the alpha peaks for the dimmer plates.}
\label{fig:VUV-LAr-comparison} 
\end{figure}
Two such tests are shown in Fig.~\ref{fig:VUV-LAr-comparison}.  The plates tested were of the same design as used in Blanche but from previous experiments such as in theTallBo dewar facility at Fermilab.  The top bar graphs show VUV monochromator measurements of two sets of wavelength shifting plates.  The relative brightness differences are apparent.  The bottom two figures show measurements of these same of plates in LAr.  The alpha peaks for the brighter plates in the VUV monochromator are significantly brighter than the alpha peaks for the dimmer plates.

\subsubsection{Wavelength Shifting Plate Performance}
\label{sect:platePerformance}

The efficiency with which wavelength shifting plates convert scintillation light  at 128~nm light to visible light can be computed as the ratio of photons read out per input photon, corrected by a geometrical factor $f_{\text{geo}}$ that accounts for the fraction of wavelength shifted light that is actually incident on the SiPM as a result of the illumination pattern of the monochromator on the sample,
\begin{equation}
\epsilon_{\text{TPB}} =  ( I^{\text{SiPM}}/I^{\text{VUV}} ) \times (\mathcal{R}^{\text{VUV}}/\mathcal{R}^{\text{SiPM}}) \times \frac{1}{f_{\text{geo}}}. 
\label{eq:efficiency}
\end{equation}
Here $( I^{\text{SiPM}}/I^{\text{VUV}} )$ is the ratio of the current read out by the SiPM divided by the current read out at 128~nm by the VUV photodiode.   Currents were measured by a Keithley 2502 dual channels picoammeter.  Errors are 0.1-0.2 pA, which are small compared with the measured currents of typically $>$10 pA.
The function $\mathcal{R}^{\text{VUV}}$ that converts the current read out by the calibrated VUV photodiode to the number of 128~nm photons illuminating the plate is given by
\begin{equation}
\mathcal{R}^{\text{VUV}} = R^{\text{VUV}}(\text{128~nm}) \times hc/\lambda(\text{128~nm}) =  2.44 \times 10^{-19}~{\text A\,s}/\gamma,
\label{eq:NISTresponse}
\end{equation}
where $R^{\text{VUV}}(\text{128~nm}) = 0.157~\text{A/W}$ is the VUV photodiode responsivity at 128~nm determined from the responsivity curve with an uncertainty estimate of 5\%.
The function $\mathcal{R}^{\text{SiPM}}$ that converts the current read out by the SiPM to visible photons needs to account for the range of photons emitted by the TPB in the plate and is given by 
\begin{equation}
\mathcal{R}^{\text{SiPM}}=\int d\lambda_{\text{vis}} \mathcal{P}_{\text{TPB}}(\lambda_{\text{vis}}) R^{\text{SiPM}}(\lambda_{\text{vis}}) \frac{hc}{\lambda_{\text{vis}}} = 4.09 \times 10^{-13}~{\text A\,s}/\gamma.
\label{eq:SiPMresponse}
\end{equation}
Here $\mathcal{P}_{\text{TPB}}(\lambda_{\text{vis}})$ is the probability that a wavelength shifted photon is produced and transmitted through the acrylic at a given visible wavelength $\lambda_{\text{vis}}$ and is shown by the solid curve in the top panel of Fig.~\ref{fig:PDsummary}.  The function $R^{\text{SiPM}}(\lambda_{\text{vis}})$ is the SiPM responsivity at $\lambda_{\text{vis}}$ that has been supplied by SensL at 5 V overvoltage.  An uncertainty of 5\% is adopted on $R^{\text{SiPM}}$.  The integral is between 250 - 600~nm because the data sets overlap in this range.

The uncertainty in the responsivity (``resp'') is assumed to be the quadratic sum of the uncertainties in the VUV and SiPM responsivities.  

A Monte Carlo simulation was used to estimate the geometrical factor $f_{\text{geo}}$.  In this simulation, 128~nm photons from the monochromator uniformly illuminate the surface of a wavelength shifting plate in the rectangular image of the VUV monochromator slit.  The dimensions of the illumination pattern were determined experimentally.  First a measurement of 128~nm light from the monochromator was made through a 4~mm machined aperture onto the 10~mm $\times$ 10~mm VUV photodiode.  Then followed a set of measurements with masks on the aperture that gave the fraction of the total light falling on its four quadrants.  From these measurements it was determined that the light from the VUV monochomator fell in a pattern on one quadrant of the wavelength shifting plate and this pattern underfilled the VUV photodiode.  The  TPB at the chosen position on the plate was then assumed to emit photons in a direction chosen from a Lambertian distribution~\cite{bib:gehman} for normally incident photons.  This simulation accounts for wavelength shifted photons that reflect off the bottom of the acrylic substrate and are typically lost.  The fraction of photons falling on the SiPM compared with the number of wavelength shifted photons emitted by the TPB was found by the Monte Carlo calculation to be $f_{\text{geo}} = 0.31$.  The imprecision in characterizing the optics of the VUV monochromator leads to an estimate of the uncertainty in $f_{\text{geo}}$ of 5\% .

Efficiencies of the four wavelength shifting plates (1-4) at 128~nm used in the Blanche experiment are given in Table~\ref{tbl:plateEff}.  These efficiencies were determined using three pieces (A, B, C) cut from each plate. 
The efficiency of each plate was computed as the mean of the 3 efficiency measurements .  
The uncertainties in the uniformity of the plate coating (``plate'') are the standard deviations divided by the means.
The uncertainty in VUV measurements (``VUV'') for plates tested in multiple trials was found to be 5.7\%.
During the measurements of the plate samples in Table~\ref{tbl:plateEff}, it was found that the response of the NIST-calibrated photodiode in the VUV monochromator had degraded and it was replaced by a new Opto Diode VUV photodiode\footnote[6]{http://optodiode.com}. 
Measurements of plate 3 after the change in photodiode showed a secular change that was unphysical when compared with measurements of plates 1,2 and 4.
A correction factor was applied to the measurements of plate 3 made earlier to account for this secular change.  
The estimated uncertainty in this correction factor was 5.9\%.

\hfil

\newpage

\begin{table}[hpt]

\begin{adjustwidth}{-1.25cm}{}

	\begin{center}
	\caption{Measurements and resulting efficiencies for the four wavelength shifting plates in Blanche}
	\label{tbl:plateEff}
	\vspace{0.2em}
	\label{tab:PlateEfficiencies}
	\begin{tabular}{| c c c || c c c |}
		\hline
		\hline
		{\small Plate/Sample} & {\small ($I^{\text{SiPM}} / I^{\text{VUV}} / 10^5 $)} & {\small $\epsilon_{\text{TPB}}$} & {\small Plate/Sample} & {\small ($I^{\text{SiPM}} / I^{\text{VUV}} / 10^5 $)} & {\small $\epsilon_{\text{TPB}}$}\\
		&  {\small [128nm]} &  &  &  {\small [128nm]} &\\
		\hline
		{\small 1/A} & {\small 3.27} & {\small 0.63}  & {\small 2/A} & {\small 3.51} & {\small 0.68} \\
		{\small ~~B} & {\small 2.87} & {\small 0.55} & {\small ~~B} & {\small 3.08} &{\small 0.59} \\
		{\small ~~C} & {\small 3.39} & {\small 0.65} & {\small ~~C} & {\small 3.51} &{\small 0.68}  \\
		\hline
		\multicolumn{3}{| l| |}{{\small$\epsilon_{\text{TPB}}$ = 0.61 $\pm$ 8.7\% (plate) }} &
		\multicolumn{3}{| l| }{{\small$\epsilon_{\text{TPB}}$ =  0.65 $\pm$ 7.4\% (plate) }} \\	
		\multicolumn{3}{| c|}{{\small $\pm$ 5.7\% (VUV) $\pm$ 7.1\% (resp) }} &
		\multicolumn{3}{| c|}{{\small $\pm$ 5.7\% (VUV) $\pm$ 7.1\% (resp) }} \\		
		\hline
		\hline
		{\small 3/A} & {\small 3.28} & {\small 0.63}  & {\small 4/A} & {\small 2.54} & {\small 0.49} \\
		{\small ~~B} & {\small 3.66} & {\small  0.71} & {\small ~~B} & {\small 2.70} &{\small 0.52} \\
		{\small ~~C} & {\small 3.29} & {\small  0.63} & {\small ~~C} & {\small 2.33} &{\small 0.45}  \\
		\hline
		\multicolumn{3}{| l| |}{{\small$\epsilon_{\text{TPB}}$ = 0.66 $\pm$ 6.3\% (plate) $\pm$ 5.7\% (VUV)}} &
		\multicolumn{3}{| l| }{{\small$\epsilon_{\text{TPB}}$ = 0.49 $\pm$ 7.2\% (plate) }} \\		
		\multicolumn{3}{| c| |}{{\small$\pm$ 7.1\% (resp) $\pm$ 5.9\% (corr)}} &
		\multicolumn{3}{| c| }{{\small $\pm$ 5.7\% (VUV) $\pm$ 7.1\% (resp) }} \\		
		\hline
		\hline
	\end{tabular}
	\end{center}

\end{adjustwidth}

\end{table}

\subsection{Light Guides}
\label{sect:lightguides}

The commercial light guides used in this experiment were manufactured by Eljen Technology\footnotemark[1].  These light guides are made from polystyrene with green-emitting EJ-280 wavelength shifter embeded in the plastic.  The absorption and emission characteristics of these light guides are shown in Fig.~\ref{fig:PDsummary}.  

In addition to their absorption and emission spectra, the performance of these light guides is characterized by their effective attenuation length in LAr.

\subsubsection{Experimental Apparatus to Measure Attenuation Length}
\label{sect:attnLenExpt}

\begin{figure}[h]
\centerline{\includegraphics[width=.3\textwidth]{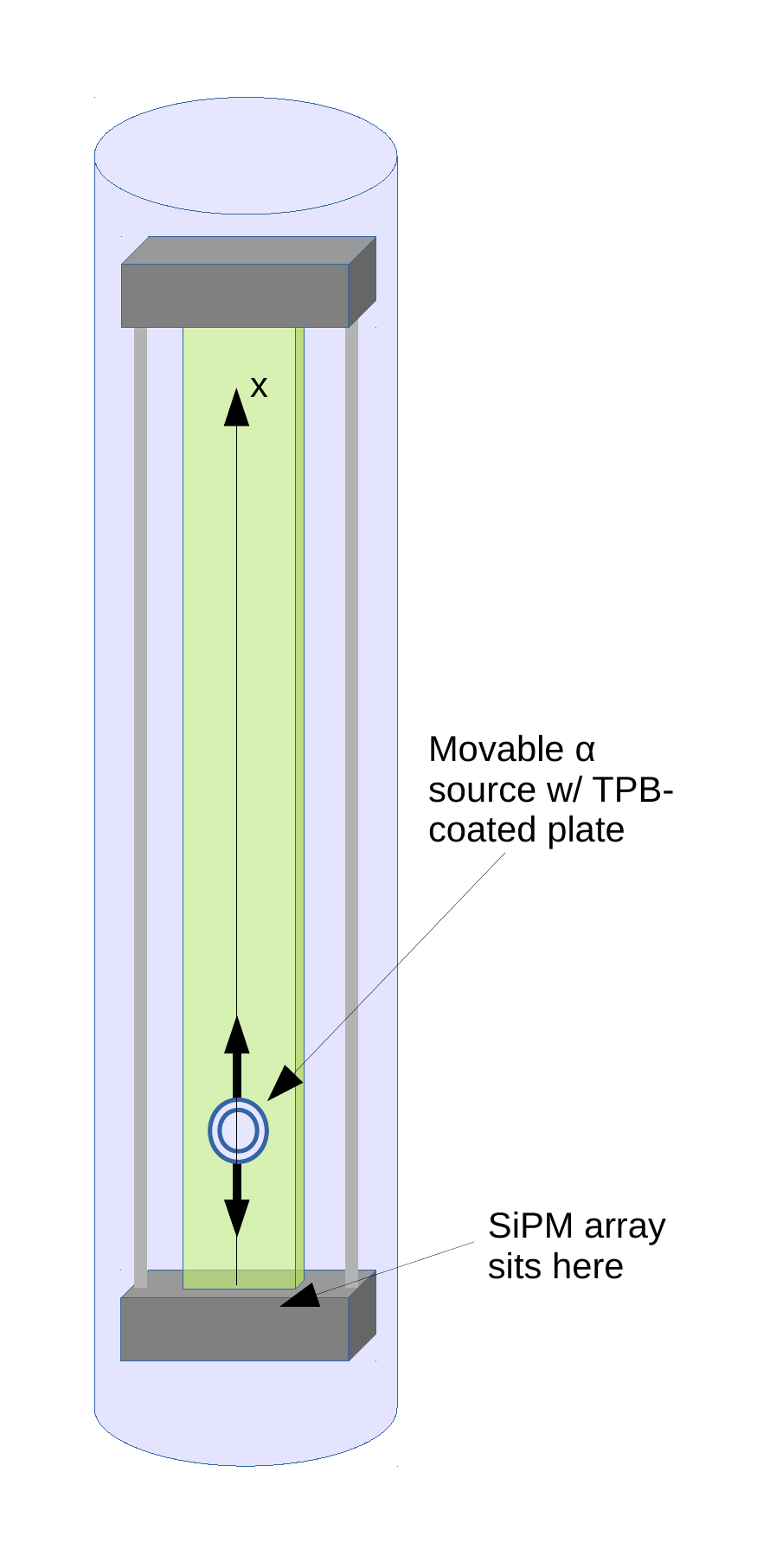}}
\caption{The apparatus used to determine the attenuation length of the Eljen light guides.}
\label{fig:attnLenApparatusCartoon} 
\end{figure}

The apparatus used at IU to measure the attenuation length of the EJ-280 light guides is depicted in Fig.~\ref{fig:attnLenApparatusCartoon}.  
The light guide is mounted in a dewar filled with LAr.   An $^{241}$Am $\alpha$ source with a $1''$ wide TPB-coated plate placed in front is fixed to a mechanism that allows it to slide along the length of the light guide.  An array of 4 SensL C-series SiPMs ($j = 0-3$) were centrally mounted in a frame fixed to the end of the light guide.  These SiPMs were read out individually as the $^{241}$Am source/TPB plate was raised through 25 positions (coordinate $x$) in 1$''$ steps.  Although the relative spacing of the $\alpha$-scan measurements are accurately determined, the absolute position of the $\alpha$ source with respect to the SiPMs is not accurately measured.  

The signals from the SiPMs were processed by a SiPM Signal Processor (SSP) module that was designed and built by the HEP Electronics Group at Argonne National Laboratory (ANL) for the DUNE photon detection system~\cite{bib:TallBo}.
The SSP includes a leading edge discriminator for the event trigger and amplitude analysis algorithms for measuring the peak and the integral of the waveform.  The SSP in this experiment was operated in ``self-triggered'' mode, in which all events generated by the $\alpha$ source that passed the threshold trigger condition led the SSP to be readout.  There were approximately 148,000 waveforms from each of the SiPMs included in the analysis.

The data analysis began by converting the integrated number of ADC counts in each event waveform at dewar position $x_i$ in SiPM $j$ 
\begin{figure}[h]
\centerline{\includegraphics[width=.85\textwidth]{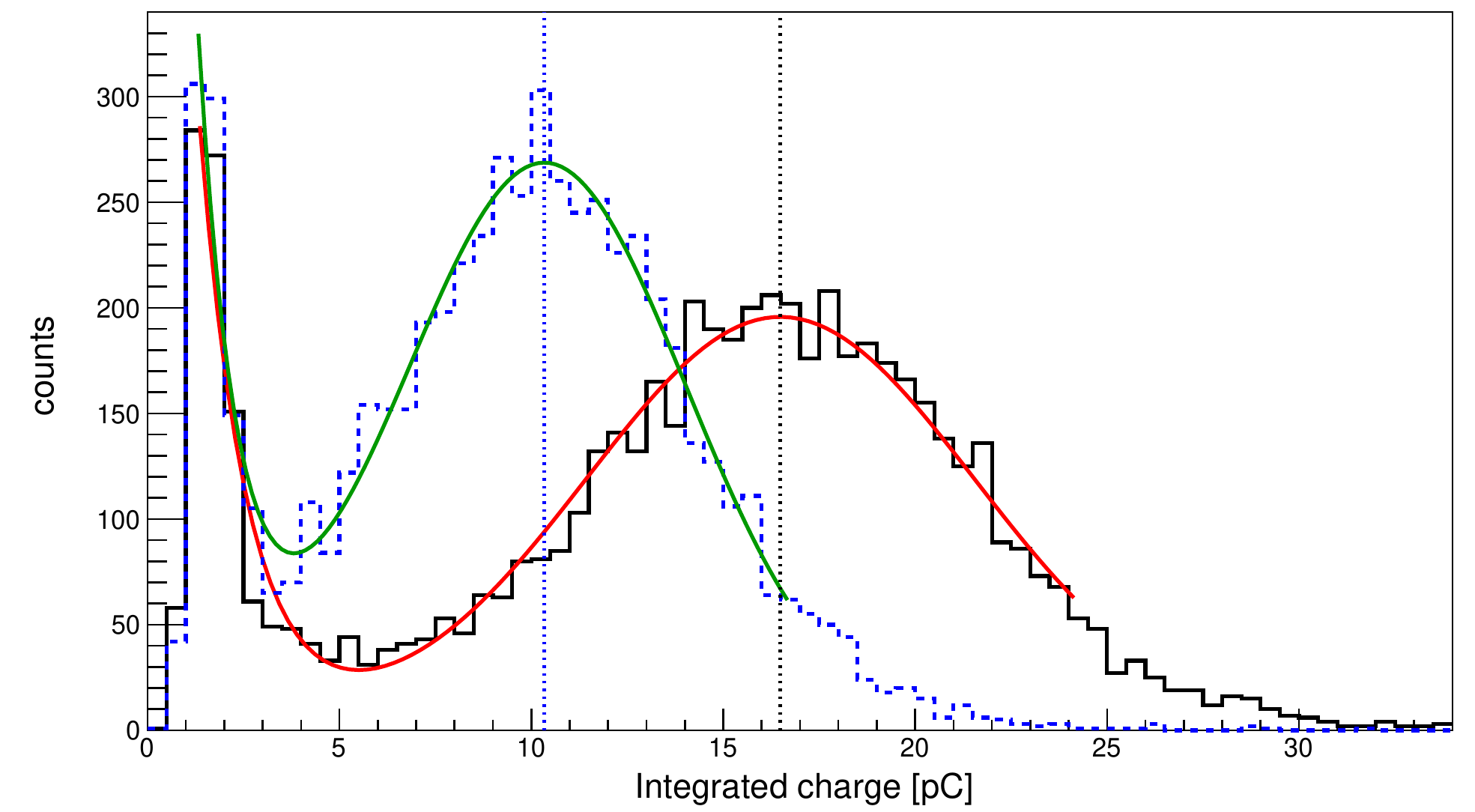}}
\caption{The $\alpha$ peak histograms for two positions $x_i$ as measured by SiPM $j = 2$. The Gaussian fits to the scintillation signals generated by the $\alpha$ source and the exponential fits to the backgrounds from SiPM dark noise or cosmics traversing the dewar are superposed.  The $\alpha$ peaks are marked by vertical lines. }
\label{fig:integratedChargeHists} 
\end{figure}
into charge, as described in~\cite{bib:TallBo}, and then collecting the integrated charge for each waveform into individual histograms $(i,j)$.  Examples of these histograms for SiPM $j = 2$ at two positions $x_i$ are shown in Fig.~\ref{fig:integratedChargeHists}.  The basic shape of these histograms has two distinct populations: a Gaussian function from the scintillation signals generated by the $\alpha$ source, and a background from SiPM dark noise or scintillation signals from cosmics traversing the dewar.  The background signals can be characterized by an exponential function.

A sum of an exponential and Gaussian function was fit to each of the ($i,j$) histograms in the analysis using ROOT\footnote[7]{https://root.cern.ch}.  The mean integrated charge from the Gaussian fits was identified as the ``$\alpha$ peak'', $p_j(x_i)$, or the peak strength of the scintillation signal at position $x_i$ for SiPM $j$.  
Fig.~\ref{fig:integratedChargeHists} shows two examples of these histogram fits with $p_j(x_i)$ marked by vertical lines.  
The  weighted means $\langle  p(x_i) \rangle$ of the 4 independent $\alpha$ peak data sets $p_j(x_i)$ were used to determine    
the attenuation length of the light guide in the Blanche experiment.  
The errors bars on $\langle  p(x_i) \rangle$ were computed as the standard deviations of the 4 $ p_j(x_i)$ used to calculate the mean.  

\subsubsection{Analysis to Determine the Light Guide Attenuation Length}
\label{sect:dataAnalysis}

The attenuation length properties of the light guides were determined by a Monte Carlo simulation that was constructed to describe the $\alpha$-scan data.  This simulation is described in \ref{lightGuideSim}.  In this simulation, the distribution of the photons from the $\alpha$ source was restricted to the center of the light guide since they were generated with a $1''$ wide TPB-coated plate adjacent to the light guide.  The generation of the photons, their capture by the waveshifter in the light guide, and their propagation down the light guide to the SiPMs at the readout end are all modeled.  In this simulation, there is one adjustable parameter, the survival probability, $\mathcal{P}$, that a photon in the light guide survives after each reflection at a wall as it propagates down the light guide.  For the simulaton used to compute the attenuation length properties, the photon survival probability at each reflection is $\mathcal{P} = 0.9988$, a parameter that describes the $\alpha$-scan data well.

\ref{lightGuideSim} describes the method used to match the $\alpha$-scan data with the simulation.  In addition to matching the $\alpha$-scan data, this method also determines the absolute position of the $\alpha$ source with respect to the SiPMs, which was not well-measured.  The data-simulation comparison results in ($\chi^2/ dof$)$_{\alpha-scan} = 17.8/25$ for the signifiance of the match.  Since this $\chi^2/ dof$ parameter characterizes the significance of the simulation match to the $\alpha$-scan data and is not the result of a fitting procedure, the number of degrees of freedom is 25, the number of data points in the $\alpha$-scan.  The absolute position of the $\alpha$ source with respect to the SiPMs in the simulation was found to be 4.83~cm, a physically reasonable result.
 
\begin{figure}[h]
\centering
\includegraphics[width=.95\textwidth]{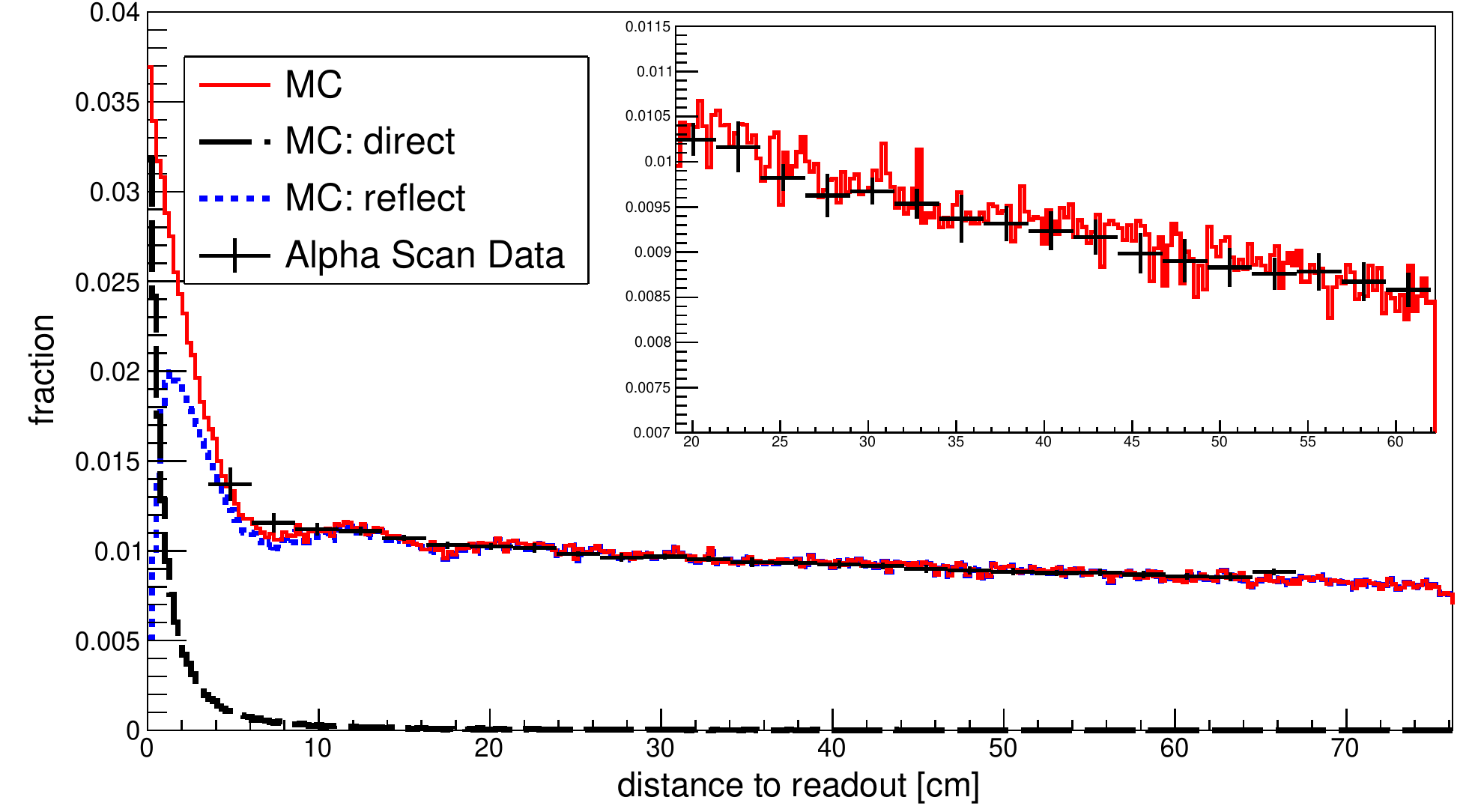}
\caption{The superposition of the $\alpha$-scan data and the simulation of the light guide.  The mean $\alpha$-scan data, $\langle  p(x_i) \rangle$, appropriately scaled, as described in  \ref{lightGuideSim}, are shown as crosses.  The fraction of photons that reach the SiPMs along direct trajectories from the wavelength shifter are shown as dashes.  The fraction of photons that reach the SiPMs along trajectories that reflect off the light guide walls are shown as dots.  The {\it inset} magnifies the midrange of the model to show the quality of the fit in that region.}
\label{fig:dataSimComparison}
\end{figure}

Fig.~\ref{fig:dataSimComparison} shows the simulation with the $\langle  p(x_i) \rangle$ from the $\alpha$-scan data superposed.  The bumps in the simulation are likely the result of artifacts of the simplified detector geometery in the simulation and are not seen at the resolution of the $\alpha$-scan data.  
In addition, Fig.~\ref{fig:dataSimComparison} shows the fraction of photons that reach the SiPMs along direct trajectories from the wavelength shifter as dashes and the fraction of photons that reach the SiPMs along trajectories that reflect off the light guide walls as dots.
The inset figure shows magnifies the midrange of the model to show the quality of the fit.  

Fig.~\ref{fig:dataSimComparison} demonstrates that the simulation procedure described in \ref{lightGuideSim} accounts for the attenuation length properties of the Eljen light guides.  Our conclusion from this data-simulation comparison is that the simulation in \ref{lightGuideSim}, suitably modified to describe the detectors in the Blanche experiment, will be effective in the analysis of the Blanche cosmic ray data. 

\subsubsection{The Transport Function for the Light Guides in the Blanche Experiment}
\label{sect:transportFunction}

\ref{techEffSim} describes the simulation used to determine the transport function for the light guides used in the Blanche experiment.  The transport function computes the probability that a photon absorbed by the wavelength shifter in the light guide reaches the readout end as a function of its starting distance from the readout end.  The simulation uses methods similar to those in \ref{lightGuideSim}.  In \ref{techEffSim} the detector parameters characterized in \S\ref{sect:sipm}, \S\ref{sect:platePerformance}, and \S\ref{sect:lightguides} are used as inputs.  The simulation assumes there are 12 SiPMs at the readout end of the light guide, as proposed for the PDs in the single-phase DUNE far detector.  As in \ref{lightGuideSim}, the photon survival probability assumed in the simulation at each reflection in the light guide is $\mathcal{P} = 0.9988$. 

A fit was made to the histogram created with this simulation model with a double exponential function of the form 
\begin{equation}
\label{eq:attnLen}
f(x) = A~e^{-x/x_{sh}} + B~e^{-x/x_{lng}},
\end{equation}
where $x$ is the distance from the readout end, $x_{sh}$ is the ``short'' attenuation length, $x_{lng}$ is the ``long'' attenuation length.   $A$ and $B$ are amplitudes that characterize the relative contribuitions of the two exponentials to the fit.  The parameters for the model fit to eq.(\ref{eq:attnLen}) using ROOT are given in Table~\ref{tab:Transport}.  In this table the normalized amplitudes, A/(A+B) and B/(A+B), are given to make more apparent the relative importance of the two exponentials.  The transport function is consequently normalized so that a photon absorbed at $x = 0$ has a probability of 1 for reaching the readout end.  The fit in Table~\ref{tab:Transport} characterizes the transport function for the Blanche detectors.  
\begin{table}[ht]
	\begin{center}
	\caption{Fit parameters for the transport function using eq.(\ref{eq:attnLen}) from simulation data created in \ref{techEffSim}}
	\vspace{0.2em}
	\label{tab:Transport}
	\begin{tabular}{| l c c c c c |}
		\hline
		\hline
		 \multicolumn{1}{|c}{}&~A/(A+B)~ &  ~~$x_{sh}$ & ~A/(A+B)~ & ~~$x_{lng}$ & ~~($\chi^2/ dof$)$_{sim}$\\
                   &     &         ~~[cm]    &  &  ~~[cm]      & \\
		\hline
		transp. func.& ~0.29 &~~4.3 & ~0.71 & ~~225 & ~~~448/296 \\
		\hline
		\hline
	\end{tabular}
	\end{center}
\end{table}

\begin{figure}[h]
\centering
\includegraphics[width=.85\textwidth]{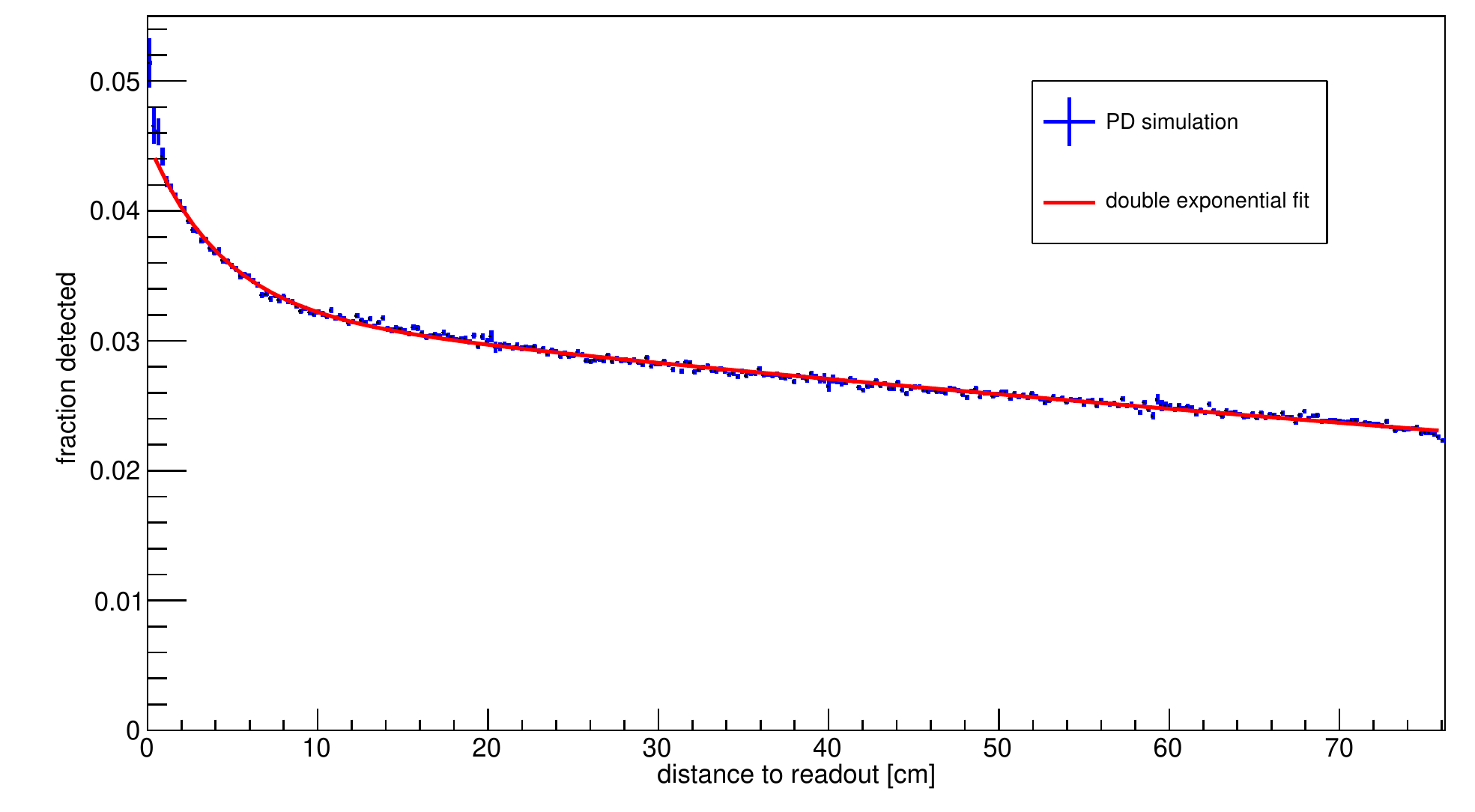}
\caption{The simulation of the transport function described in \ref{techEffSim} for the detectors in the Blanche experiment with the double exponential model fit in Table~\ref{tab:Transport} superposed. }
\label{fig:fullSim}
\end{figure}

The simulation of the transport function with the double exponential model overlayed is shown in Fig.~\ref{fig:fullSim}.  In this figure, the amplitudes are not normalized.  
The large value of $\chi^2/ dof = 1.5$ for this fit results from two effects.  First, only the small statistical uncertainties from a large statistics simulation are used in the calculation of $\chi^2$ by the fitter.  In addition, a double exponential model is only a simplification of the way light falls off from the readout end of the light guide.  A more complicated model, however, is not justified by the experimental data and this simplified model is assumed to be adequate to determine the absolute efficiency of this detector technology.

\subsection{Efficiency of the PD Technology from Laboratory Measurements}
\label{sect:calcPdEfficiency} 

\begin{figure}[h]
\centering
\includegraphics[width=.95\textwidth]{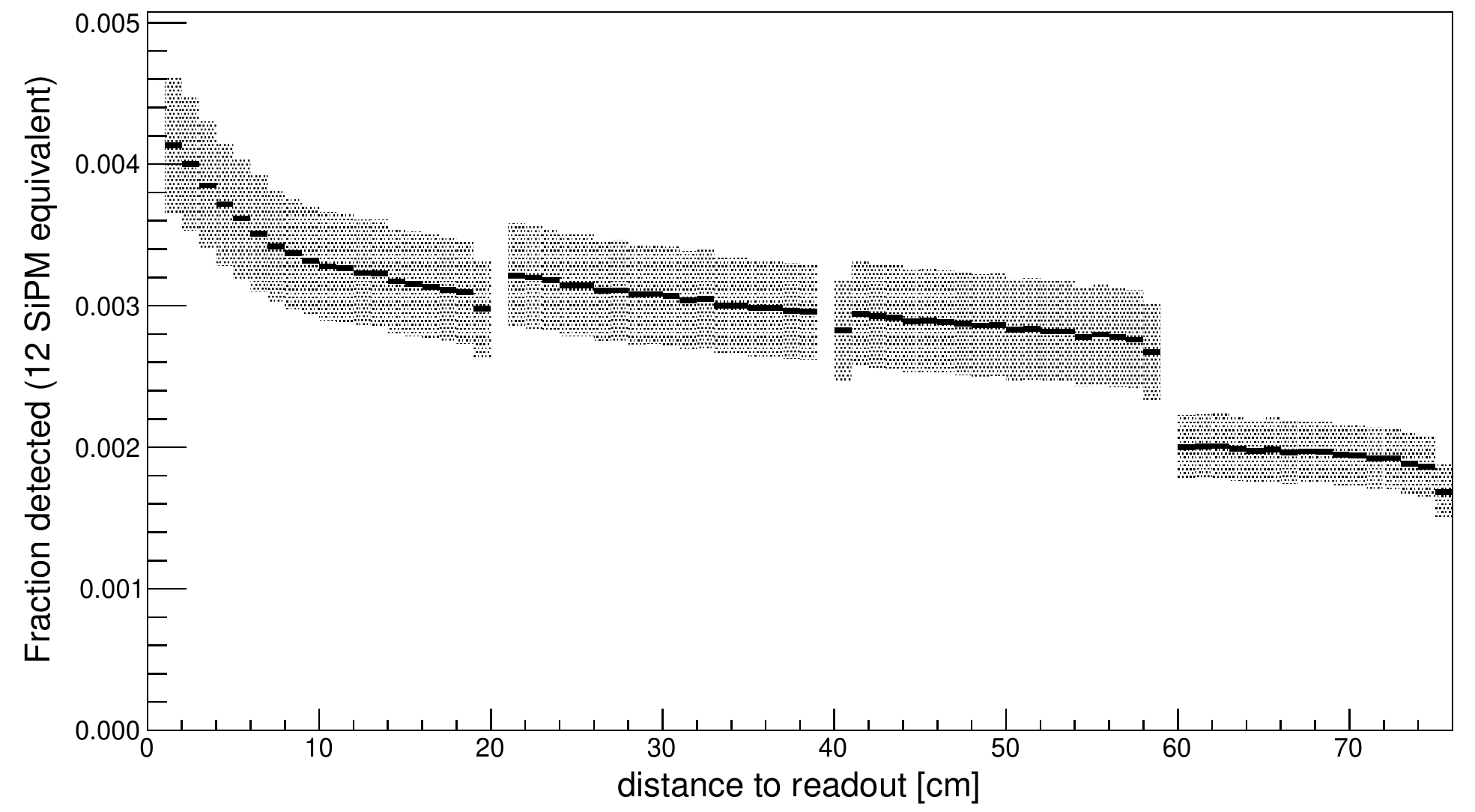}
\caption{Simulation of the fraction of photons detected by the SiPMs as a function of starting position along the length of the photon detector, or the efficiency of this PD technology, as determined from laboratory measurements and simulations.  The four separate pieces of the efficiency results are due to the different efficiencies (1 - 4 from the left) for the wavelength shifting plates in Table~\ref{tbl:plateEff}.  The gaps in the efficiency are due to the gaps in the plates, as illustrated in Fig.~\ref{fig:PDcartoon}.}
\label{calculatedEfficiency}
\end{figure}

The efficiency of the PD technology described in \S\ref{sec:PhotonDetectorDesign} has also been computed using the simulation in \ref{techEffSim}.  
The simulation makes a histogram that gives the fraction of photons detected by the PD technology as a function of their starting positions from the SiPM readout.  This histogram is shown in Fig.~\ref{calculatedEfficiency}.  The physical extent of the plate 1 does not extend to $x=0$, so the first point shown is at $x=1.5$~cm.
The systematic errors shown in the figure are calculated from Table~\ref{tab:PlateEfficiencies} and also the error in $f_{geo}$.
This figure should be interpreted as the efficiency of this PD technology as a function of position along its length.  The four separate sections of the efficiency histogram are due to the different efficiencies of the wavelength shifting plates in Table~\ref{tbl:plateEff}. The uncertainties in the efficiency are dominated by the errors in the plate efficiencies.

Owing to the different efficiencies of the wavelength shifting plates, it is difficult to simply characterize the efficiency of this photon detector technology from Fig.~\ref{calculatedEfficiency}.  The approach taken to estimating the efficiency of the technology tested in the laboratory was to rerun the simulation in \ref{techEffSim} with the mean plate efficiency for plates 1, 2, and 3 in Table~\ref{tbl:plateEff}.  Plate 4 was dropped from the average as one that would be identified as unacceptable for the DUNE detector.  The detector parameters in \S\ref{sect:sipm} and \S\ref{sect:lightguides} were left unchanged.  The resulting histogram model was then fit with the double exponential function transport function multiplied by a costant efficiency parameter.  At the readout end ($x = 0$), the absolute efficiency was found to be 0.48\%.  

The results of these laboratory measurements lead to the absolute efficiency of this photon detector technology given in eq.(\ref{eq:eff}),
\begin{equation}
\epsilon = 4.8 \times 10^{-3} \times [0.29 \exp{(-x/4.3) } + 0.71 \exp{(-x/225)}],
\label{eq:eff}
\end{equation}
where $x$ is measured in $cm$.  At $x = 0 $ cm, the efficiency is $4.8 \times 10^{-3}$.  In eq.(\ref{eq:eff}), $x$ is measured in $cm$.  When averaged over the length of the 76.2 cm light guide, the efficiency is $3.0 \times 10^{-3}$.

\section{Experimental Test of Photon Detector Design at Blanche}
\label{sect:Blanche}

\FloatBarrier
\subsection{Experimental Design}
\label{sec:BlancheExperimentalDesign}

\begin{figure}[h]
\centering
\includegraphics[width=.45\textwidth]{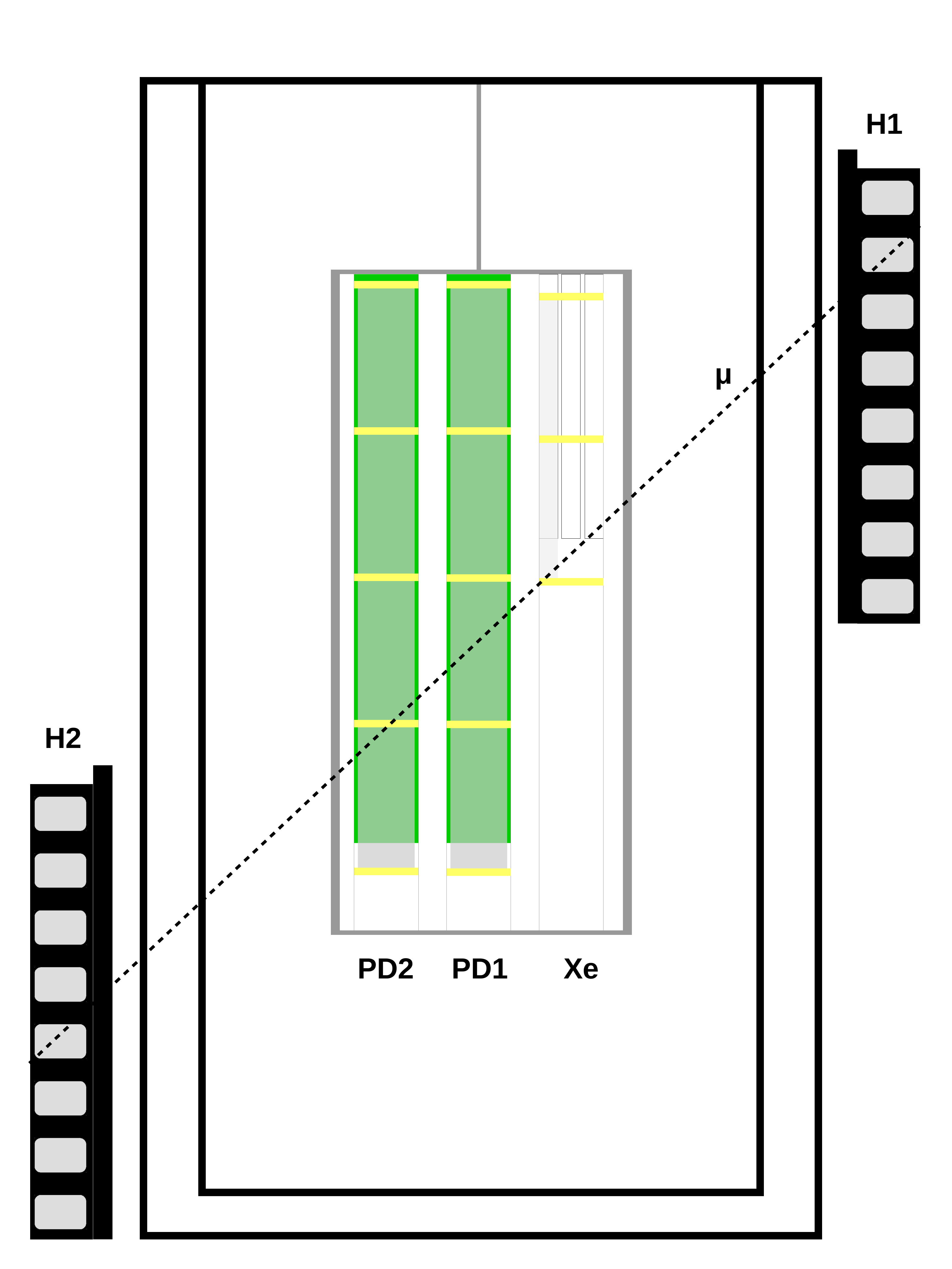}
\caption{A drawing of the experimental apparatus in the Blanche dewar. Two photon detector modules, PD1 and PD2, are mounted in a custom frame suspended from the dewar lid.  A set of detectors for a separate experiment, labeled Xe, are also mounted in the frame.  Coincidence triggers that assure events pass through the detectors are provided by the adjacent hodoscope trigger paddles H1 and H2, which flank the outside of the dewar.  A representative triggered cosmic muon trajectory is superposed.}
\label{blancheExpt}
\end{figure}

The experiment took place in the liquid argon dewar facility ``Blanche'' at the Proton Assembly Building (PAB) at Fermi National Accelerator Laboratory (FNAL).  The experiment ran from September 14, 2016 until September 27, 2016.  
Fig.~\ref{blancheExpt} shows a schematic of the experiment.  
The experimental apparatus consisted of two prototype DUNE photon detector modules (PD1 and PD2), as pictured in Fig.~\ref{fig:PDcartoon}, on a custom paddle mount suspended from the lid.  The front and back sides of both PD1 and PD2 had 4 waveshfitng plates mounted above their surfaces.  

Two hodoscope modules were installed on opposite sides of the Blanche dewar to select single-track cosmic-ray muons passing through the LAr volume.  Each hodoscope module consists of 64 2-inch diameter barium-fluoride crystals, coated with TPB and arranged in an 8$\times$8 array. Each crystal is monitored by a PMT. 
Since the hodoscope modules were originally designed to detect bremsstrahlung photons in the CREST balloon flight experiment 
they are very sensitive to extraneous photon activity around our experiment. To reject this $\gamma$ ray activity, a pair of plastic scintillator panels covering the entire face of a hodoscope module were placed between each hodoscope module and the Blanche dewar. These panels were individually read out by PMTs. The SSP readout was triggered by four-fold coincidence logic that required at least one hit in each hodoscope module on opposite sides of the dewar, as well as one hit in their two adjacent scintillator planes. Typically this trigger indicates that each event contains at least one charged particle passing through the liquid argon.  Events were further filtered offline to reject showers by requiring one and only one hit in each hodoscope module. Single-track events crossing from one side of the frame to the other were rejected in order to exclude any tracks that could pass through a light guide.

There were four data sets collected in the experiment: PD1 front side, PD1 back side, PD2 front side, and PD2 back side.  
Only the data from the front side of PD1 were used in the analysis.  Data from the back side of PD1 were excluded due to mounting failures during and after installation, which have the effect of reducing contributions from reflected light in the dewar.  Data from both sides of PD2 were excluded for multiple reasons.  Most important, the attenuation length of the light guide in PD2, determined by $\alpha$-scan after the experiment had ended, was less than $\sim$1~m.  Since the primary objective of this experiment was to determine the absolute efficiency of the IU PD technology, the PD2 data would not give accurate results for a photon detector that would be used in DUNE.

\FloatBarrier
\subsection{Blanche Operations}
\label{sec:operations}

To prepare for a run, the Blanche dewar was first evacuated by a turbo pump to help reduce contamination from residual gases and then back-filled with gaseous argon.  The gaseous argon was next replaced with ultra-high-purity (UHP) LAr that passed through a molecular sieve and copper filter that had been regenerated just prior to the run.  The volume of LAr in Blanche was approximately 570 liters. 

The contaminants that most affect LAr scintillation light are O$_2$, N$_2$, and H$_2$O, which can both quench scintillation light and decrease the argon transparency at 128~nm~\cite{bib:MITN2,bib:O2Contamination,bib:O2contamCross,bib:H2O,bib:H2OcontamCross}.  UHP LAr is typically delivered with low levels of these contaminants and the regenerated filters are very effective at further removing them.  

The O$_2$, N$_2$, and H$_2$O concentrations were monitored during the run.  
The O$_2$ contamination stabilized at $\sim$30~ppb when sampled for a week at the end of the run.   At concentrations $<$100~ppb, the effects of O$_2$ contamination on scintillation light in LAr are negligible~\cite{bib:O2Contamination}.  
The N$_2$ contamination was $\sim$80--90~ppb.  It was sampled at the begining of the run and then for a week at the end of the run.  At these concentrations, the N$_2$ does not affect the scintillation light~\cite{bib:MITN2,bib:N2Contamination}.  
The H$_2$O contamination was measured to be $\sim$5~ppb or less.  It was sampled for 6 days at the beginning of the run.  It was again sampled approximately 10 days after the end of this experiment (for a second experiment with the same apparatus) and the concentration was again $\sim$5~ppb or less.
The effects of H$_2$O contamination on LAr scintillation light are not well studied.  
However, studies in gaseous argon suggest the H$_2$O concentration at this level does not affect our results~\cite{bib:H2O}.

Once filling was complete, Blanche was then sealed and subsequently maintained at a positive internal pressure of 8 psig to ensure no contamination from outside during the run.  Gaseous argon from the ullage was recondensed to liquid argon with a liquid nitrogen condenser and then returned to the dewar. 

\subsection{Analysis of Blanche Cosmic Ray Data}
\label{sec:BlancheAnalysis}

The data set analyzed for the Blanche experiment is made up of tracks that cross the front side of PD1 and tracks that meet both the four-fold trigger criterion and the requirement of only one PMT hit in each hodoscope module.  Two additional selection cuts were made.  (1) To exclude spurious triggers with no activity in the dewar, PD1 and PD2 were each required to collect at least 10 photoelectrons (PE) from the track.   (2) To exclude triggers from high energy events unlikely to be minimum ionizing muons, events with relatively high energy deposition in the hodoscope crystals were removed.  There were 11,223 tracks analyzed that passed the cuts.

The waveforms collected by each SiPM in the track were pedestal subtracted and summed from 0.7~$\mu$s before the trigger until 11~$\mu$s after the trigger.  The result is 
\begin{figure}[h]
\centering
\includegraphics[width=.85\textwidth]{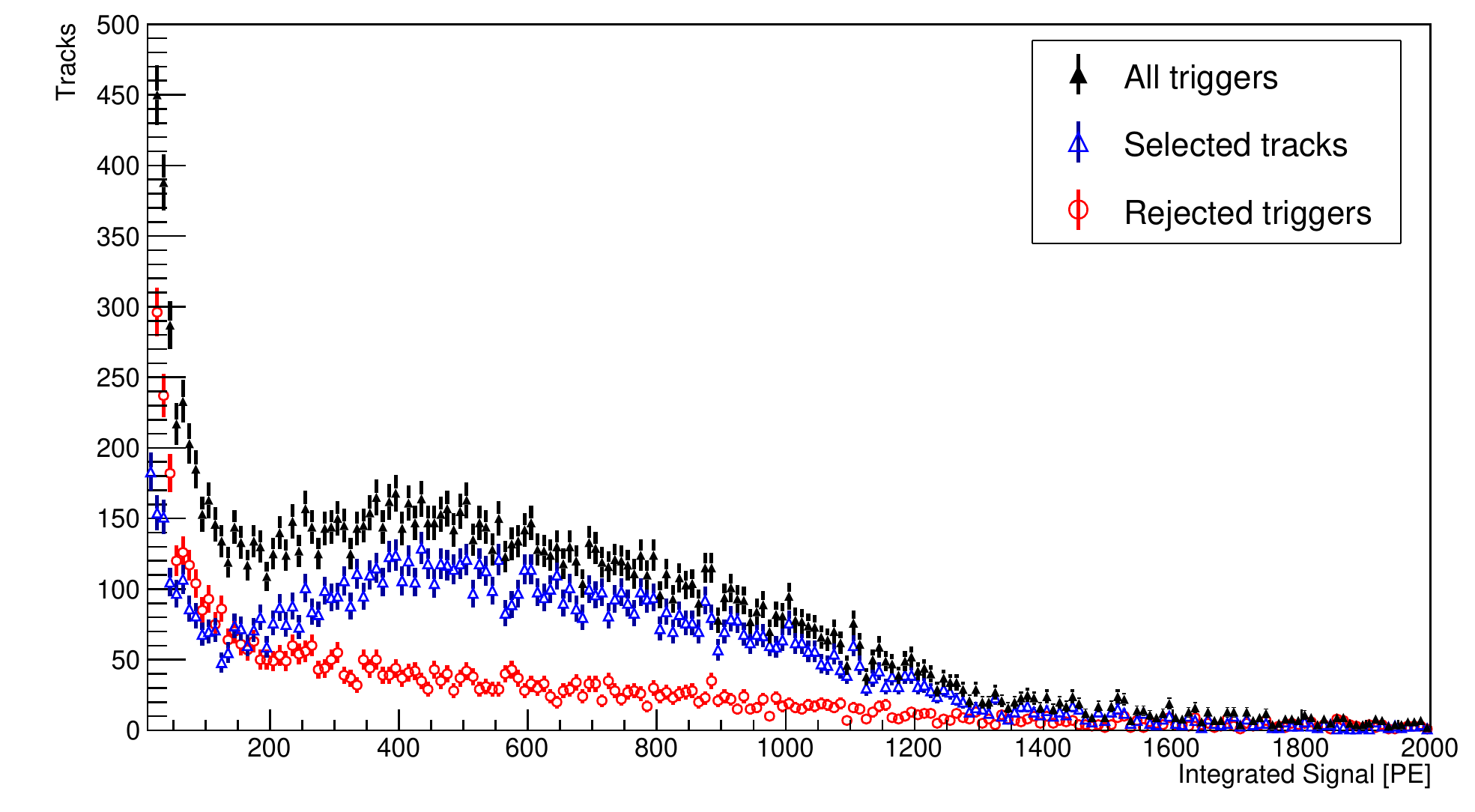}
\caption{The histograms of the integrated PE from all preselected track-like triggers.  Also shown are the histograms for the tracks passing all cuts and the triggers excluded from the data set.  The "Selected tracks" are the cosmic muon data set analyzed. }
\label{fig:blanche-PEdetected}
\end{figure}
the total number of ADC counts in the track for each SiPM.  For the calibration of ADC counts to PE, a data set was recorded in the ``self-triggered mode'', in which all signals above threshold were recorded with minimal cuts, and a histogram was filled that showed the separation between the single PE peaks.  This separation provides the PE calibration.  The calibrated PE counts in each SiPM were then summed.  Fig.~\ref{fig:blanche-PEdetected} shows a histogram of the integrated PE from all triggers.  
Also shown in the figure are the histograms for the tracks passing all cuts and the triggers excluded from the data set.  The integrated number of PE for the selected tracks, $N_{det}$, makes up the cosmic muon data set selected for analysis.  

For each selected track, the expected number of scintillation photons falling on the surface of PD1 as a function of distance from the 8 SiPMs on the readout end was simulated as described in  \ref{trackSim}.  
The histogram of photons from each track was then attenuated bin by bin by the photon transport function in Table~\ref{tab:Transport}.  The result of integrating the histogram for each track yields $N_{exp}$, the expected number of scintillation photons at the readout end from that track.  The ratio of $N_{det}/N_{exp}$ was then computed for each track in the sample.  

There are multiple uncertainties in $N_{det}/N_{exp}$.  There is a calibration uncertainty associated with converting the measured ADC counts in the track to photoelectrons.  This calibraton constant was determined by averaging the separations of the single PE peaks individually for the 8 SiPMs in the experiment.  Characterizing the uncertainty for each SiPM as standard deviation divided by the means for the used peak separations and averaging these 8 values yields 3.5\%.  The second uncertainty is the statistical uncertainty associated with the simulation of $N_{exp}$.  As described in \ref{trackSim}, this error is small, typically $\sim$0.2\% and no larger than 0.45\%.  
 
\begin{figure}[h]
\centering
\includegraphics[width=.85\textwidth]{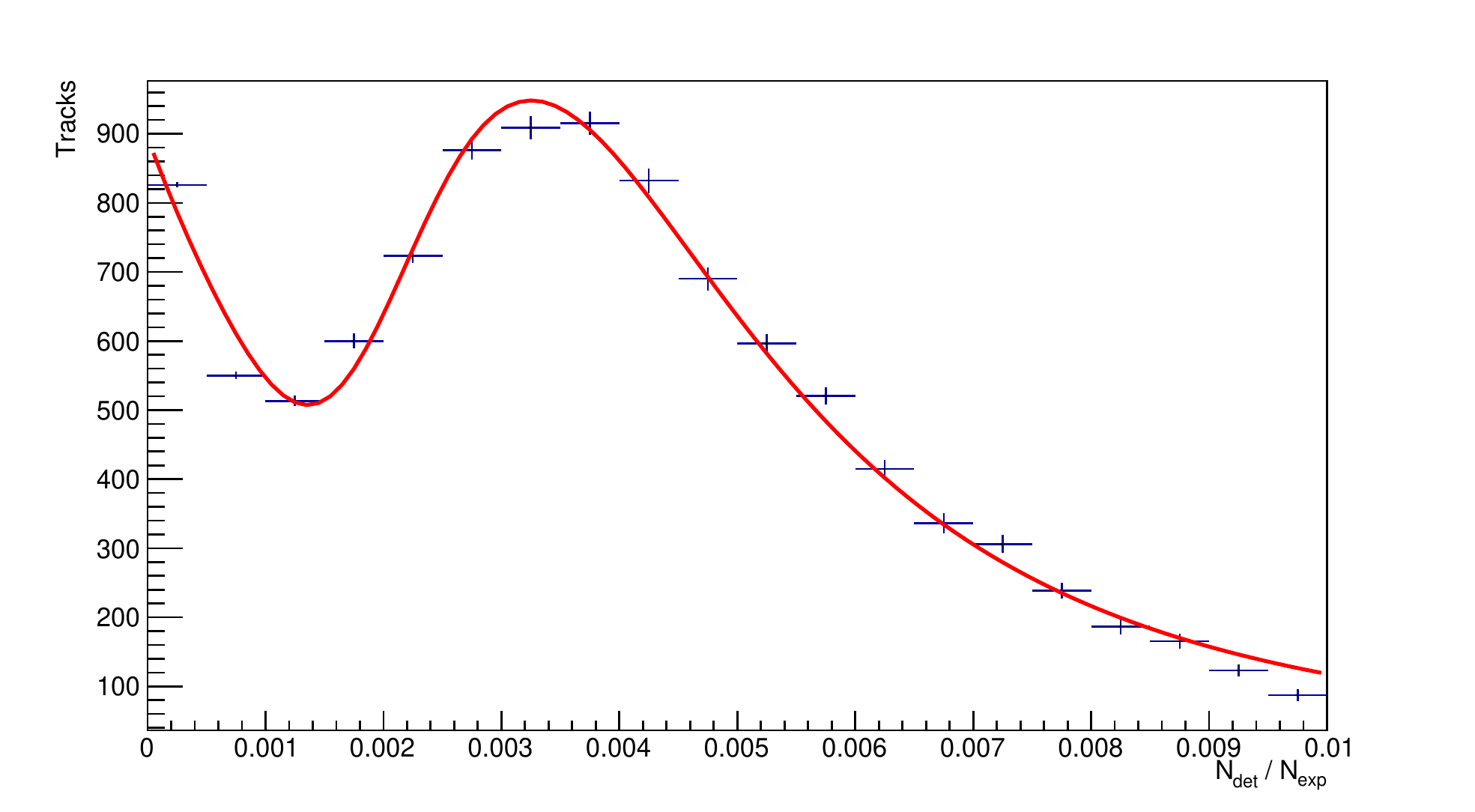}
\caption{The ratio of the number of detected PE to the number of photons expected, $N_{det}/N_{exp}$ averaged over 100 trials, for all tracks in the selected data sample in Fig.~\ref{fig:blanche-PEdetected} are shown with error bars.  Superposed is the average model as described in the text.  The most probable value of $N_{det}/N_{exp}$ is 0.0035 $\pm$ 0.3\% PE/photon.}
\label{fig:PEperPhoton-withFit}
\end{figure}

A Monte Carlo procedure was adopted to find the most probable value (MPV) for the number of expected photoelectrons per incident photon for the tracks in the Blanche data set.  For each track in the sample, 100 values of $N_{det}/N_{exp}$ were generated with normally distributed random errors derived from the quadratic sum of the calibration error and the error in simulating $N_{exp}$.  These 100 values for each track were placed into 100 histograms.  The average of the 100 histograms of $N_{det}/N_{exp}$ is shown in Fig.~\ref{fig:PEperPhoton-withFit}.  
The errors on $N_{det}/N_{exp}$ is the standard deviation of the values in the histogram bin.  

The 100 individual histograms were each fit using ROOT with a model that superposes an exponential background with a Landau function.  The MPV for each of the 100 model histograms was determined by finding the peak of the fitted Landau distribution.  The average of the 100 values of the MPV is 0.0035 $\pm$ 0.3\% PE/photon, where the error comes from the standard deviation divided by the average.  The average of the fit parameters for the 100 models was used to generate an ``average'' model.  This average model is superposed on average $N_{det}/N_{exp}$ histogram in  Fig.~\ref{fig:PEperPhoton-withFit}.

In the current baseline design for the DUNE far detector, 12 SiPMs are planned for the readout end of the light guide.  This would lead to an increase in $N_{det}/N_{exp}$ for this technology by 12/8 to 0.0053 PE/photon.

The correction to $N_{det}/N_{exp}$ for the cross talk probabilities in Table~\ref{tab:SiPM-Noise} was taken from the prescription in~\cite{bib:GallegoXtalk}, modified for the properties of the SensL SiPMs.  A Monte Carlo calculation for the probability of cross talk adding between 1 and 4 additional counts for a trigger initially firing a single microcell gives an average of 1.25 -- 1.26 additional counts per detected photon for multiple models.  Cross talk consequently reduces $N_{det}/N_{exp}$ to 0.0041 photons detected per incident photon.  
Although the prescription in~\cite{bib:GallegoXtalk} covers only up to 4 additional cross talk counts for a single microcell hit, this range accounts for $> 99.7\%$ of all cross talk counts and the calculation was considered adequate.  No correction was made for afterpulsing.

\subsection{Efficiency of the PD Technology from the Blanche Experiment}
\label{sect:BlanchePdEfficiency} 

\begin{figure}[h]
\centering
\includegraphics[width=.85\textwidth]{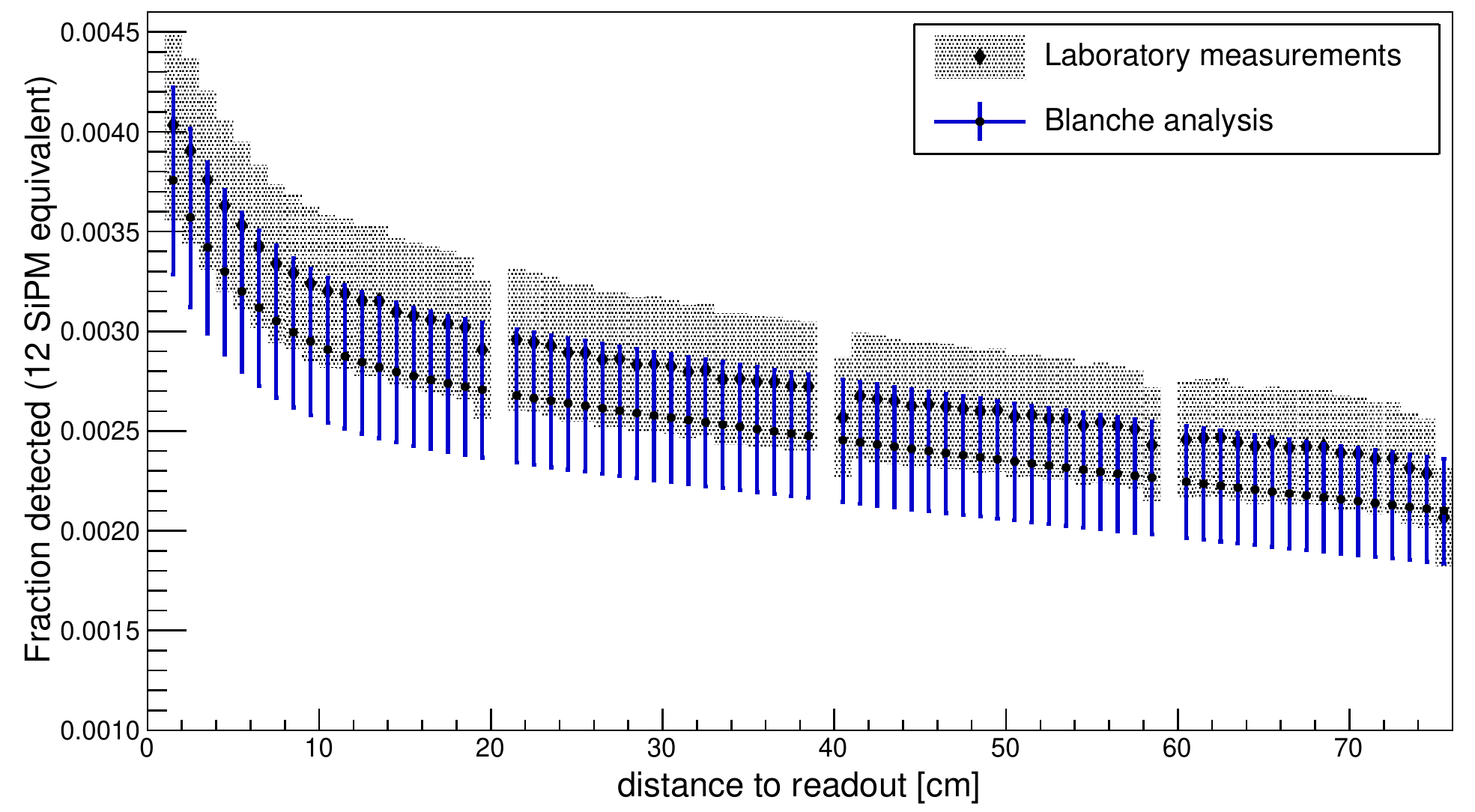}
\caption{The comparison of the absolute efficiency as a function of distance from the readout end for the PD technology developed at Indiana University for the single phase far detector of DUNE.  Determinations from both laboratory measurements and the Blanche cosmic ray analysis are shown.  In this figure the light yield of minimum ionizing cosmic muons as they traverse LAr is ${\mathcal N}_{phot} = 40,000$ photons/MeV~\cite{bib:Miyajima,bib:Doke1,bib:Doke2,bib:scintYield2}.}
\label{fig:compEfficiency}
\end{figure}

The absolute efficiency of the PD technology from the  Blanche experiment is defined as $N_{det}/N_{exp}$ for 12 SiPMs as a function of distance from the SiPM readout after correcting for cross talk.  At the readout end ($x=0$), the efficiency is given by $\epsilon = 4.1 \times 10^{-3}$.  The absolute efficiency of PD technology, with the transport function in Table~\ref{tab:Transport}, is shown in Fig.~\ref{fig:compEfficiency}.  
The gaps, including the gap at the readout end, correspond to positions not covered by wavelength shifting plates.  The systematic errors on the Blanche analysis shown in the figure are calculated from Table~\ref{tab:SystematicUncertBlanche}.
\begin{table}[h]
  \begin{center}
    \caption{Systematic uncertainties in the Blanche analysis}
    \vspace{0.2em}
    \label{tab:SystematicUncertBlanche}
    \begin{tabular}{|  c  |  c  |}
      \hline
      \hline
      Systematic & ~~~ Uncertainty ~~~ \\
      \hline
      \hline
      Scintillation yield~\cite{bib:Miyajima,bib:Doke1,bib:Doke2,bib:scintYield2} & $\pm$10\% \\
      \hline
      Rayleigh scattering~\cite{{bib:RayleighScat},bib:rayleighSeidel, bib:rayleighIshida, bib:RayleighScatteringNeumeier12} & +2.3\%, -3.1\% \\
     \hline
      MPV & $\pm$0.3\% \\
     \hline
      VUV reflections, stainless steel& $\pm$7\% \\
      \hline
      \hline
   \end{tabular}
  \end{center}
\end{table}

Overlaid on this figure is the efficiency determined from laboratory measurements as a function of position from the readout end.  For this figure the laboratory efficiency was modified from Fig.~\ref{calculatedEfficiency} by re-running the simulation in \ref{techEffSim} with the wavelength shifting plates having equal efficiencies instead of those given in Table~\ref{tbl:plateEff}.  
Equal efficiencies were used because scintillation photons from muon tracks typically illuminate all the plates, not just one as in the $\alpha$ scan.  The monolithic efficiency for the plates used in the \ref{techEffSim} simulation was calculated as the weighted average of the efficiencies in Table~\ref{tbl:plateEff}, with the weights set by the number of photons falling on the plates from the sample tracks computed in \ref{trackSim}.  
\hfil
\newpage

\section{Discussion}
\label{sect:discussion}

This paper reports the characterization of a prototype VUV photon detection technology for DUNE made up of wavelength shifting plates and commercially manufactured light guides that are read out by SiPMs.   Two studies were made.    
As described in \S\ref{sect:calcPdEfficiency}, one study measured the characteristics of individual components in the lab and then used these measurements in a simulation of a photon detector module to determine the photon detection efficiency as a function of distance from the readout end. 
As described in \S\ref{sec:BlancheAnalysis}, a second study was made using through-going cosmic ray muon tracks in LAr with an integrated prototype photon detector module in Blanche.
Comparing detected photons from the tracks against simulated photon exposure attenuated with the light guide transport function provides a second measurement of the efficiency of this technology.
Clearly these two efficiency determinations must match.  The comparison shown in Fig.~\ref{fig:compEfficiency} suggests these two determinations are in reasonable agreement.  

There is only one adjustable parameter available in these efficiency determinations -- the light yield of minimum ionizing cosmic muons as they traverse LAr, ${\mathcal N}_{phot}$~photons/MeV, which is used in the analysis of the Blanche data.
In \ref{trackSim}, ${\mathcal N}_{phot}$ parameterizes $N_{exp}$, the number of scintillation photons expected on the surface of PD1, which in turn is used to compute the efficiency, $N_{det}/N_{exp}$, from the Blanche data.
In Fig.~\ref{fig:compEfficiency}, minimum ionizing muons are assumed to generate ${\mathcal N}_{phot} = 40,000$ photons/MeV as they cross the LAr volume~\cite{bib:Miyajima,bib:Doke1,bib:Doke2,bib:scintYield2}.  
In the efficiency calculation based on laboratory measurements shown in Fig.~\ref{fig:compEfficiency}, there are no adjustable parameters.  

The overlapping error bars in Fig.~\ref{fig:compEfficiency} strongly suggest that the two efficiency determinations are consistent, which 
supports the assumption made in the Blanche analysis that minimum ionizing muons generate ${\mathcal N}_{phot} = 40,000$ photons/MeV as they cross the LAr volume.

\newpage

\noindent {\bf Acknowledgements} \\

\noindent This work was supported in part by the Trustees of Indiana University, the DOE Office of High Energy Physics through grant DE-SC0010120 to Indiana University, and grant \#240296 from Broookhaven National Laboratory to Indiana University.  The authors wish to thank the many people who helped make this work possible. At IU: M.~Gebhard, M.~Lang, B.~Martin, J.~Musser, P.~Smith, J.~Urheim.  At Fermilab: R.~Davis, C.~Escobar, B.~Miner, E.~Niner, S.~Pordes, B.~Rebel, F.~Schwartz, M.~Zuckerbrot.  At ANL: J.~Anderson, G.~Drake, A.~Kreps, M.~Oberling.  At Colorado State:  N.~Buchanan, J.~Jablonski, D.~Warner. At Eljen Technology: C. Hurlbut. 

This manuscript has been authored by Fermi Research Alliance, LLC under Contract No. DE-AC02-07CH11359 with the U.S. Department of Energy, Office of Science, Office of High Energy Physics. The United States Government retains and the publisher, by accepting the article for publication, acknowledges that the United States Government retains a non-exclusive, paid-up, irrevocable, world-wide license to publish or reproduce the published form of this manuscript, or allow others to do so, for United States Government purposes.

\hfil

\newpage

\bibliographystyle{elsarticle-num}

\bibliography{IUplateTechnology}



\vspace{0.5cm}

\appendix
\section{$\alpha$-Scan Simulation}
\label{lightGuideSim}

The $\alpha$-scan simulation is based on a model detector with components whose dimensions and compositions are described in the main text.  The simulation modeled the enitre 76.2~cm length of the light guide (Fig.~\ref{fig:PDcartoon}).  

The $\alpha$-scan simulation starts by distributing 128~nm scintillation photons from the $\alpha$ source onto a 1$''$ wide wavelength shifting plate embedded with TPB that runs the length of the light guide and is separated from it by 0.5$''$.  This photon distribution mimics how the scintillation photons induced by the $\alpha$ source illuminate the light guide in the apparatus in Fig.~\ref{fig:attnLenApparatusCartoon}.  The wavelengths of the resulting wavelength shifted photons are chosen from the TPB emission spectrum determined in the lab by reflectance measurements; the photons' directions are chosen from a Lambertian distribution for normally incident photons.  If a photon's direction is away from the light guide, that photon is considered lost and the simulation generates a new photon.  If a photon is directed toward the light guide, its transmission probability through the $\frac{1}{16} ''$ acrylic plate determines whether it reaches the wavelength shifting plate boundary.  The index of refraction for the acrylic is 1.49.  If a photon reaches the boundary, its angle of incidence determines whether it is reflected back into the wavelength shifting plate by total internal reflection at the acrylic-LAr interface and lost or whether it exits the plate.  The index of refraction for LAr is 1.23.

For photons exiting the wavelength shifting plate, their trajectories are recalculated using Snell's law and their intersection points with a plane at the top surface of the polystyrene light guide are computed.  If the intersection point falls on the light guide, the photons' trajectories are again recalculated with Snell's law at the LAr - polystyrene boundary, using an index of refraction for the Eljen polystyrene of 1.6.  Otherwise, the photon is lost.  The simulation then determines whether the photons hitting the light guide are absorbed by the wavelength shifter in it.  Using the EJ-280 absorption coefficient measurements provided by Eljen and an exponential deviate drawn from a function in ROOT's TRandom3 package, an absorption length for each photon is calculated.  If this absorption length is longer than the distance to the next interface with a light guide surface, the photon is lost.  If shorter, the photon is captured by the wavelength shifter.  As reported by Eljen, the efficiency for emission is $86\% \pm 1.7\%$.  If a uniformly chosen random number is $\leq$0.86, a new photon is generated with a wavelength drawn from the emission spectrum shown in the bottom panel of Fig.~\ref{fig:PDsummary} and a direction uniform in solid angle.  Photons are required to have wavelengths equal or longer than the photons absorbed.  

Photons are next tracked through the lightguide.  Along their trajectories they either reflect off a light guide surface if their angle of incidence is greater than the critical angle or they are lost.  At each reflection there is also a survival probability, $\mathcal{P}$, that photons survive and are {\it not} lost at each reflection due to some (unspecified) physical process.  In this simulation, $\mathcal{P}$ is the one adjustable parameter.   Photons are tracked until they reach one of the four centrally mounted SiPMs on the readout end of the light guide or they are lost.  Since there is overlap in the EJ-280 absorption and emission spectra, reabsorptions/re-emissions are permitted along the trajectory.  Photons that reach the readout end are considered detected if they meet two criteria.  First, the photon must strike at the physical position of one of the four SiPMs.  Second, the photon must be ``seen'' by the SiPM, in the sense that a random number drawn for that photon is less than the SiPM PDE ({\it c.f.}, bottom panel of Fig.~\ref{fig:PDsummary}) for that wavelength.  Photons are lost if they do not meet these criteria.  

For each detected photon, a histogram is filled with the position where it was initially absorbed by the wavelength shifter in the light guide.  For every photon generated by the simulation absorbed by the light guide, a second histogram is filled with the position where it was initially absorbed by the wavelength shifter.  After simulating $\sim5\times10^8$ photons, the histogram filled with the detected photons was divided by the histogram where all the photons started.  The resulting histogram gives the fraction of photons detected as a function of their distance in the light guide from the readout end, $\mathcal{F}(x_i)$.  An example histogram is shown in Fig.~\ref{fig:dataSimComparison} for the light guide used in the Blanche experiment.  In addition, separate histograms were filled with those photons that propagate directly to the SiPMs and those that bounce off the walls before reaching the SiPMs.  These additional histograms are also shown in Fig.~\ref{fig:dataSimComparison}.

Simulation models were compared with the $\langle p(x_i) \rangle$ data from the $\alpha$-scans.  The simulation model with $\mathcal{P} = 0.9988$ was found to give a reasonable $\chi^2$/d.o.f. $< 1$ for the fit.  There was no attempt to make  $\mathcal{P}$ more precise by searching though parameter space with multiple simulations since the accuracy of the $\alpha$-scan data do not justify small changes in $\mathcal{P}$ for marginal improvements in $\chi^2$/d.o.f.  For this comparison the $\mathcal{F}(x_i)$ were averaged into 1$''$ bins to match the spacing of the $\alpha$-scan data.  In addition, a nominal value  for the $\alpha$ source position was chosen.  After averaging and choosing the $\alpha$ source position, the $\mathcal{F}(x_i)$ were integrated over the range of the $\alpha$-scan data and the area under the simulation curve, $A_{sim}$, determined.  Next the $\langle p(x_i) \rangle$ data were integrated over their range and the area under the $\alpha$-scan curve, $A_\alpha$, was determined.  The units of $A_\alpha$ are [pC].  The shapes of the simulation curve and the $\alpha$-scan curve can now be directly compared by normalizing the $\alpha$-scan curve by $A_{sim}/A_\alpha$. 

For this choice of the $\alpha$ source position, the simulation was compared with the $\alpha$-scan data $\langle p(x_i) \rangle$.  The comparison was characterized by a $\chi^2$ parameter given by 
\begin{equation}
\chi^2 = \sum_i \frac{[\langle  p(x_i) \rangle \times (A_{sim}/A_\alpha)- \mathcal{F}(x_i)]^2}{e^2(x_i)\times (A_{sim}/A_\alpha)^2},
\label{eq:chisqTot}
\end{equation}
where $e(x_i)$ is the standard deviation of the 4 $ p_j(x_i)$ used to compute $\langle  p(x_i) \rangle$.  It is important to recognize that the $\chi^2$ parameter in eq.(\ref{eq:chisqTot}) is being used to characterize how well the 25 $\alpha$-scan data points and the choice of the $\alpha$ source position match the simulation, not to fit a model.  For this reason, the number of d.o.f. = 25.

The absolute position of the $\alpha$ source with respect to the SiPMs was then varied and the fitting procedure repeated.  At a separation of the $\alpha$ source from the SiPMs of $x = 4.83$~cm, a physically reasonable result, $\chi^2$/d.o.f. =17.8/25 was at its minimum value.

\vspace{0.5cm}

\section{Simulation of the PD Technology Efficiency from Laboratory Measurements}
\label{techEffSim}

The Monte Carlo simulation for the efficiency of the PD technology from the laboratory measurements in \S\ref{sect:sipm}, \S\ref{sect:platePerformance}, and \S\ref{sect:lightguides} is similar to the simulation used to find the attenuation length of light guides in \ref{lightGuideSim}.  VUV scintillation photons strike a plate embedded with TPB where they are wavelength shifted into the visible; the wavelength shifted photons are then propagated to the light guide where they can be trapped and wavelength shifted again; in the final step of the simulation the photons are transported through the light guide to the SiPMs where they are tested for detection.  This simulation explicitly uses the efficiency of the plates from Table~\ref{tbl:plateEff} and the photon survival probability at each reflection $\mathcal{P} = 0.9988$ determined in \S\ref{sect:lightguides} to compute the efficiency.  

The simulation starts by distributing 128~nm scintillation photons from LAr uniformly onto the four wavelength shifting plates on the light guide, as shown in Fig.~\ref{fig:PDcartoon}.  The wavelengths and directions of the wavelength shifted photons are chosen as in \ref{lightGuideSim}.  The transmission of the photons through the plate and into the light guide, and their absorption by the light guide are simulated as in \ref{lightGuideSim}.  Photons are tracked through the light guide with the probability that a photon is lost at each reflection set at $\mathcal{P} = 0.9988$.  Photons that reach the readout end are detected based on a random number drawn and tested against the SiPM PDE in the bottom panel of Fig.~\ref{fig:PDsummary}.  

Approximately $10^9$ photons were simulated and two histograms were created.  
One histogram was filled with the position where each detected photon was initially absorbed by the wavelength shifter in the light guide and a second histogram was filled with the starting position of every simulated photon that was absorbed by the wavelength shifter in the light guide.  These two histograms were divided to give a histogram with the fraction of photons detected as a function of their starting positions in the light guide.  This histogram was fit with the double exponential function eq.(\ref{eq:attnLen}) to give the transport function for the PD technology.  The fit results are reported in Table~\ref{tab:Transport}. 

\begin{figure}[h]
\centering
\includegraphics[width=.80\textwidth]{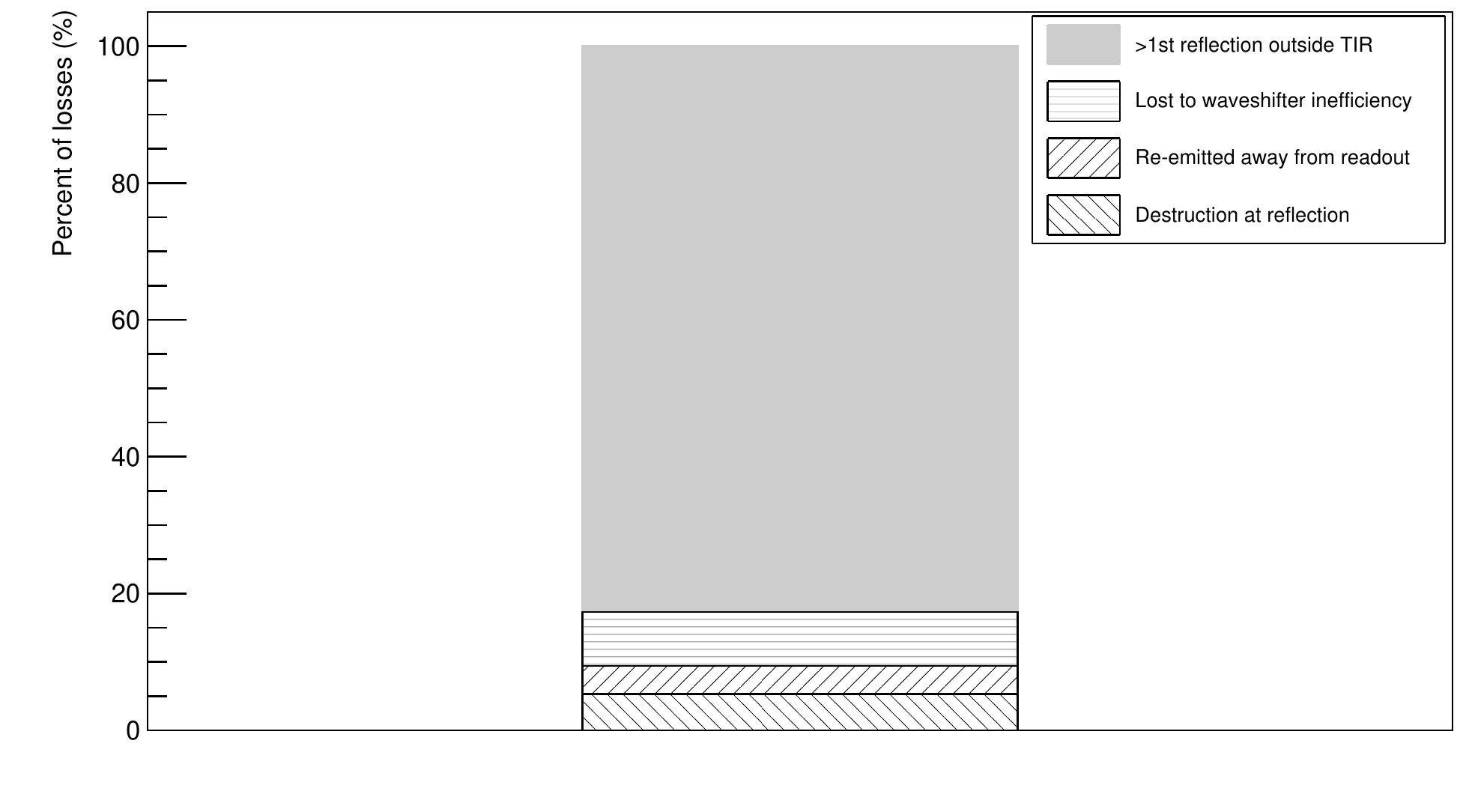}
\caption{The realtive contribution of processes that account for the loss of photons as they travel towards the readout end.  (1) Photons destroyed at a boundary with a probability of $1-\mathcal{P} = 0.0012$.  (2) Photons lost to EJ-280 wavelength shifter inefficiency.  (3) Photons lost when they are reabsorbed by the wavelength shifter and reemitted in a direction away from the readout end.  (4) Photons lost when their first reflection off a boundary is greater than the critical angle for total internal reflection but subsequent reflections are at less than the critical angle.}
\label{fig:lostPhotons}
\end{figure}

There are four processes that account for the loss of photons as they travel towards the readout end: (1) photons can be destroyed at a boundary with a probability of $1-\mathcal{P} = 0.0012$; (2) photons lost to EJ-280 wavelength shifter inefficiency;(3) photons can be lost when they are reabsorbed by the wavelength shifter and reemitted in a direction away from the readout end; and (4) photons can be lost when their first reflection off a boundary is greater than the critical angle for total internal reflection but subsequent reflections are less than the critical angle.  Fig.~\ref{fig:lostPhotons} shows the relative contribution of these three processes to the loss of photons initially traveling towards the readout.  
This figure schematically shows that most photons are lost as they move toward the readout end when their first reflection off a boundary is greater than the critical angle but then subsequently reflect off a boundary at less than the critical angle.

A second set of histograms was also created from the simulated photons.  
The first was filled with the starting position of each detected photon on the wavelength shifting plate.  Two weights were assigned to each histogram entry.  One weight accounts for the fractional coverage of the SiPM photosensitive area at the end of the light guide.  For 12 (6x6)mm$^2$ SiPMs, this weight is 0.84.  The second weight accounts for the efficiency of the plate on which the photon started.  This weight was taken from Table~\ref{tbl:plateEff}.  
The second histogram was filled with the starting position of every simulated photon on the wavelength shifting plate. 
These two histograms were then divided to give a histogram with the fraction of photons detected as a function of starting position on the wavelength shifting plate, or the efficiency of the PD technology from laboratory measurements that is shown in Fig.~\ref{calculatedEfficiency}.

\vspace{0.5cm}

\section{Simulation of Cosmic Muon Tracks in Blanche}
\label{trackSim}

For the track simulation through Blanche, cosmic muons are treated as traveling along straight paths with end points fixed at the centers of the two triggered PMTs in the hodoscopes on either side of the dewar.  It is assumed in the simulation that cosmic muons are minimum ionizing particles with an energy deposition in LAr of 2.105~MeV/cm ~\cite{bib:PDG}.  Let ${\mathcal N}_{phot}$ be the number of scintillation photons per MeV muons emit along their tracks.  Then the number of photons emitted by muons per cm along their tracks = $2.105 \times {\mathcal N}_{phot}$~photons/cm.  Consequently, the length of a muon track in the LAr volume determines how many scintillation photons are generated.

Typically ${\mathcal N}_{phot}$ = 40,000 photons/MeV~\cite{bib:Miyajima,bib:Doke1,bib:Doke2,bib:scintYield2} is used in simulations of photon production in LAr and was also chosen here.  In this simulation, ${\mathcal N}_{phot}$ is the only adjustable parameter available to match the efficiency of the PD technology as determined by the Blanche experiment with the efficiency of the technology as determined from laboratory measurements in \S\ref{sect:calcPdEfficiency} .

The starting positions of the scintillation photons were distributed uniformly along the track segment passing through the LAr volume and the photons' momentum vectors were distributed uniformly in solid angle.  Photons were tracked along straight line paths until they intersected with the photon detector, the dewar wall or were lost.  Photons could be lost if they were absorbed by a contaminant or the dewar wall, or if they hit the top of the liquid level.  Along their track, photons could also undergo a Rayleigh scattering or reflect off the walls of the dewar.  

From its starting point, each photon's trajectory was first calculated to its intersection point with the PD or the dewar wall.  The simulation then determines whether the photons are absorbed by a contaminant.  The probability that a scintillation photon is absorbed by a contaminant along its path is $P_{abs} = A \exp{(-L/\lambda_{abs})}$, where $A$ is a normalization constant, $L$ is the track length to the intersection, and $\lambda_{abs}$ is the absoption length.   In this expression
\begin{equation}
\lambda_{abs} = 1/n_c\sigma = 1/[\chi \cdot n_{LAr} \cdot \sigma],
\end{equation}
where $n_c =$ the number density of contaminants, $\sigma$ is the cross section for absorption of 128~nm VUV photons, $\chi =$ fractional number of contaminant molecules, and $ n_{LAr}$ = the number density of LAr atoms/cm$^3$  = (1.396~g/cm$^3$)/(39.948~g/mol) $\times \, (6.02 \times 10^{23}$~atoms/mol)~\cite{bib:PDG} $ = 2.1037\times 10^{22}$ LAr atoms/cm$^3$.  Table~\ref{tab:contaminants} gives $\lambda_{abs}$ for the contaminants measured in the Blanche experiment.
\begin{table}[ht]
	\begin{center}
	\caption{Absorption length for contaminants.}
	\vspace{0.2em}
	\label{tab:contaminants}
	\begin{tabular}{| c c c c c |}
		\hline
		\hline
		 contaminant&  ~~~~~$\chi$~~~ &  ~~~~~$\sigma$~~~ & ~~~~$\lambda_{abs}$~~~& Ref.\\
                       &     ~~[ppb]        & ~~[cm$^2$] &  ~~[m]     &  \\
		\hline
		 N$_2$ &~~85 & ~$7.1 \times 10^{-21}$ & ~~783  &\cite{bib:MITN2}\\
		 O$_2$ &~~30 & ~$2.8 \times 10^{-19}$ & ~~56.7 & \cite{bib:O2contamCross}\\
		 H$_2$O &~~5 & ~$8.0 \times 10^{-18}$ & ~~11.9  &\cite{bib:H2OcontamCross}\\
		\hline
		\hline
	\end{tabular}
	\end{center}
\end{table}

An exponential deviate drawn from ROOT's TRandom3 package and $\lambda_{abs}$ from Table~\ref{tab:contaminants} were then used to calculate an absorption length for the photon.  If this absorption length is shorter than the distance (keeping track of its total path length) to the intersection, the photon is lost.  

The probability that the photon is Rayleigh scattered along its path is given by $P_{Rayleigh} = B \exp{(-L/\lambda_{Rayleigh})}$, where $B$ is a normalization constant, $L$ is the track length to the intersection, and $\lambda_{Rayleigh} = 1.1$~m~\cite{bib:RayleighScat} is the Rayleigh scattering length.  There are many different values for $\lambda_{Rayleigh}$ found in the literature\cite{bib:rayleighSeidel, bib:rayleighIshida, bib:RayleighScatteringNeumeier12}.  The Rayleigh scattering length used here is a recent estimate that falls between the upper and lower values.  An exponential deviate drawn from ROOT's TRandom3 package and $\lambda_{Rayleigh}$ were then used to calculate a Rayleigh scattering length for the photon.  

The Rayleigh scattering process is handled in the simulation with the same procedure as the G4OpRayleigh code in Geant4\footnote[8]{http://geant4.web.cern.ch/geant4/} (version 4.10.03), but with the Geant4 specific three-vector and random number commands coverted to ROOT commands.  For this code, the simulated photons are assumed to be unpolarized.  The scattered photon's direction and polarization angle are assigned by the code and its energy is unchanged.  After scattering, a new Rayleigh scattering length was chosen and an intersection point was computed for the trajectory, which was tested for absorption or Rayleigh scattering.  The systematic errors reported in Table~\ref{tab:SystematicUncertBlanche} were determined by rerunning the simulation with $\lambda_{Rayleigh} = 0.66$~m~\cite{bib:rayleighIshida} and $\lambda_{Rayleigh} = 1.63$~m~\cite{bib:RayleighScatteringNeumeier12}.

Photons can also be reflected off or be absorbed by the walls of the dewar.  The reflection of 128~nm scintillation photons off stainless steel in LAr is not well studied.  In this simulation, the probability of reflection is assumed to be 25\%, equally split between specular and diffuse reflection \cite{bib:IcarusReflection}.  For specular reflections, the outgoing photon's direction was computed with respect to the normal to the wall and the rotation of the polarization vector in the perpendicular plane was unchanged.  For diffuse reflections, the outgoing photon's direction was chosen to be uniform in solid angle and its polarization vector was chosen to uniform in the perpendicular plane.  The systematic uncertainty associated with the reflectivity was estimated by simulation.  Variations in the most probable value for the MPV in Fig.~\ref{fig:PEperPhoton-withFit} were computed for the probability of reflection ranging from 12.5\% and 37.5\%.  In addition, the most probable value for the MPV was computed for the reflections being 100\% specular and 100\% diffuse.  These studies lead to a 7\% estimate for the systematic uncertainty associated with the reflectivity.

Every track in the Blanche data set that struck two single hodoscope PMTs and passed in front of PD1 was simulated 10 times.  For each of these 10 simulated tracks, the positions of all photons that reach the PD module were collected into a histogram.  The resulting 10 histograms were then averaged to determine the expected number of photons that reach the readout end, $N_{exp}$, for all muon tracks in the Blanche data set.   The fractional uncertainty in $N_{exp}$, as determined from the standard deviation of the 10 simulations using the nominal Rayleigh scattering length is typically $\sim$0.2\%.  No uncertainties are greater than 0.45\%.

\hfill

\end{document}